\def\submissioncopy{1}
\documentclass[aps,prx,preprint,notitlepage,superscriptaddress,nofootinbib,floatfix]{revtex4-2}

\usepackage{amsmath,amssymb,bm}
\usepackage{graphicx}
\usepackage{booktabs}
\usepackage{placeins}
\usepackage{xcolor}
\usepackage{hyperref}

\hypersetup{
  colorlinks=true,
  linkcolor=blue,
  citecolor=blue,
  urlcolor=blue,
  pdftitle={Transformer Atomic Cluster Expansion: TRACE},
  pdfauthor={Paramvir Ahlawat}
}
\renewcommand{\arraystretch}{1.15}

\newcommand{\calN}{\mathcal{N}}
\newcommand{\calL}{\mathcal{L}}
\newcommand{\vect}[1]{\bm{#1}}

\newif\ifshowrevisions
\ifdefined\submissioncopy
  \showrevisionsfalse
\else
  \showrevisionstrue
\fi
\definecolor{draftfivecolor}{RGB}{0,92,155}
\newcommand{\draftfive}[1]{%
  \ifshowrevisions\textcolor{draftfivecolor}{#1}\else#1\fi}
\definecolor{draftsevencolor}{RGB}{137,48,85}
\newcommand{\draftseven}[1]{%
  \ifshowrevisions\textcolor{draftsevencolor}{#1}\else#1\fi}
\definecolor{draftninecolor}{RGB}{0,105,92}

\definecolor{drafttencolor}{RGB}{166,72,0}

\definecolor{draftelevencolor}{RGB}{94,53,177}
\newcommand{\drafteleven}[1]{%
  \ifshowrevisions\textcolor{draftelevencolor}{#1}\else#1\fi}
\definecolor{drafttwelvecolor}{RGB}{183,28,28}
\newcommand{\drafttwelve}[1]{%
  \ifshowrevisions\textcolor{drafttwelvecolor}{#1}\else#1\fi}
\definecolor{draftthirteencolor}{RGB}{46,125,50}
\newcommand{\draftthirteen}[1]{%
  \ifshowrevisions\textcolor{draftthirteencolor}{#1}\else#1\fi}
\definecolor{draftfourteencolor}{RGB}{21,101,192}
\newcommand{\draftfourteen}[1]{%
  \ifshowrevisions\textcolor{draftfourteencolor}{#1}\else#1\fi}
\definecolor{draftfifteencolor}{RGB}{194,24,91}

\definecolor{draftsixteencolor}{RGB}{109,76,65}

\definecolor{draftseventeencolor}{RGB}{0,77,64}

\definecolor{drafteighteencolor}{RGB}{63,81,181}

\definecolor{draftnineteencolor}{RGB}{198,40,40}
\newcommand{\draftnineteen}[1]{%
  \ifshowrevisions\textcolor{draftnineteencolor}{#1}\else#1\fi}
\definecolor{drafttwentycolor}{RGB}{2,119,189}
\newcommand{\drafttwenty}[1]{%
  \ifshowrevisions\textcolor{drafttwentycolor}{#1}\else#1\fi}
\definecolor{drafttwentytwocolor}{RGB}{123,31,162}

\definecolor{drafttwentythreecolor}{RGB}{0,121,107}

\definecolor{drafttwentyfourcolor}{RGB}{216,67,21}
\newcommand{\drafttwentyfour}[1]{%
  \ifshowrevisions\textcolor{drafttwentyfourcolor}{#1}\else#1\fi}
\definecolor{drafttwentyfivecolor}{RGB}{142,36,170}
\newcommand{\drafttwentyfive}[1]{%
  \ifshowrevisions\textcolor{drafttwentyfivecolor}{#1}\else#1\fi}

\definecolor{drafttwentysixcolor}{RGB}{85,107,47}
\newcommand{\drafttwentysix}[1]{%
  \ifshowrevisions\textcolor{drafttwentysixcolor}{#1}\else#1\fi}
\newcommand{\drafttwentysixeq}[1]{%
  \ifshowrevisions{\color{drafttwentysixcolor}#1}\else#1\fi}
\definecolor{drafttwentysevencolor}{RGB}{0,105,120}

\definecolor{drafttwentyeightcolor}{RGB}{191,54,12}
\newcommand{\drafttwentyeight}[1]{%
  \ifshowrevisions\textcolor{drafttwentyeightcolor}{#1}\else#1\fi}
\definecolor{draftthirtycolor}{RGB}{0,110,135}
\newcommand{\draftthirty}[1]{%
  \ifshowrevisions\textcolor{draftthirtycolor}{#1}\else#1\fi}

\definecolor{draftthirtyonecolor}{RGB}{126,45,119}

\newenvironment{draftthirtyoneblock}
  {\ifshowrevisions\color{draftthirtyonecolor}\fi}
  {}
\definecolor{draftthirtytwocolor}{RGB}{0,107,83}
\newcommand{\draftthirtytwo}[1]{%
  \ifshowrevisions\textcolor{draftthirtytwocolor}{#1}\else#1\fi}
\newenvironment{draftthirtytwoblock}
  {\ifshowrevisions\color{draftthirtytwocolor}\fi}
  {}

\begin{document}
\title{Transformer Atomic Cluster Expansion: TRACE}
\author{Paramvir Ahlawat}
\email{paramvir.chem@gmail.com}
\noaffiliation

\begin{abstract}
Designing machine-learning interatomic potentials involves achieving the precise representation of complex many-body interactions alongside the efficiency required for scalable molecular dynamics. We introduce Transformer Atomic Cluster Expansion (TRACE), an energy-conserving architecture that combines atomic cluster expansion density correlations with local multihead cross-attention. The correlations form an O(3)-equivariant state for each center, which queries tensorial neighbor features that remain fixed functions of species and geometry. No learned state is passed between atoms. On a laptop MacBook-M1, we train and test TRACE for polymorphic cesium lead iodide, liquid water, and intramolecular methyl migration against experiments. For cesium lead iodide, TRACE reproduces the r$^2$SCAN+rVV10 ordering of four polymorphs and gives a classical edge-sharing hexagonal non-perovskite($\delta$) to corner-sharing cubic perovskite($\alpha$) Gibbs-free-energy crossing $\simeq$580K near the experimental observations of $\simeq$600K. By employing enhanced sampling to cross high energy barriers, the same TRACE potential successfully captures the $\delta$-to-$\alpha$ perovskite transformation without any reinforcement learning. A water potential trained on a reduced set of CCSD(T) configurations places the first oxygen--oxygen maximum at 2.85~\AA, compared to the experimental value of 2.80~\AA{}. For the gas-phase methyl migration in 2,2-dimethylisoindene, umbrella sampling yields an activation free energy of $27.92\pm0.03$~kcal~mol$^{-1}$, in close agreement with the experimental measurement of $29.2\pm1.1$~kcal~mol$^{-1}$. Across these diverse benchmarks, a single unified architecture successfully captures multi-species crystallization, liquid structures, phase diagrams, and chemical reactivity.
%For gas-phase methyl migration in 2,2-dimethylisoindene, umbrella-sampling give an activation free energy of $27.92\pm0.03$~kcal~mol$^{-1}$, compared with experimentally measured $29.2\pm1.1$~kcal~mol$^{-1}$. Across these tests, the same architecture describes phase diagrams and crystallization of multi-species crystals, liquid structure, and a chemical reaction.
\end{abstract}

% Retain RevTeX preprint typography without placing the abstract on a new page.
\makeatletter
\@booleanfalse\preprintsty@sw
\makeatother
\maketitle
\makeatletter
\@booleantrue\preprintsty@sw
\makeatother

% \section*{Popular Summary}
%

\newpage

\section{Introduction}
Molecular dynamics is inherently linear in time. Because atoms vibrate so rapidly, advancing a trajectory requires constant step-by-step force calculations at extremely small time increments. Although quantum mechanics provides the most accurate picture of chemical bonding, applying it at every single step demands massive computational power. Consequently, highly accurate simulations remain restricted to small systems and short time scales. Empirical force fields reach larger scales by imposing a chosen functional form. Machine-learning interatomic potentials (MLIPs) instead learn the Born--Oppenheimer potential-energy surface from electronic-structure data and evaluate it at a much lower cost \cite{behler2007,bartok2010,thompson2015,shapeev2016}.

A useful MLIP should represent both the geometry and the symmetries of an atomic environment. The energy is unchanged by translation, rotation, inversion, or permutation of equivalent atoms, whereas internal vector and tensor features must transform with the corresponding coordinates. The energy must also vary smoothly~\cite{pota2024thermal} because forces, stress, phonons, elastic response, and molecular dynamics depend on its derivatives. These requirements are commonly combined with a local decomposition
\begin{equation}
 E(\mathcal R,\vect h)=\sum_{i=1}^{N}E_i(\mathcal X_i),
 \label{eq:local-energy-intro}
\end{equation}
where $\mathcal R=\{(\vect r_i,Z_i)\}_{i=1}^{N}$ is the atomic
configuration, $\vect h$ is the periodic cell when present, and $\mathcal X_i$ contains the neighbors of atom $i$ within a finite cutoff. For a bounded neighbor count, Eq.~\eqref{eq:local-energy-intro} has linear cost in the number of atoms. The primary remaining challenge is to maximize the expressivity of each local energy $E_i$ while maintaining computational efficiency. 

The atomic cluster expansion (ACE) provides a rigorous foundation for describing local many-body geometry. By expanding the neighbor density into radial functions and spherical harmonics, ACE systematically couples angular momenta to construct scalar and tensor correlations~\cite{drautz2019}. Equivariant neural networks also rely on the representation theory, and learning nonlinear maps between irreducible tensor channels~\cite{geiger2022}. The integration of these features into graph machine learning architectures marked a major leap in performance; notably, NequIP introduced equivariant message passing for interatomic potentials and demonstrated exceptional data efficiency~\cite{batzner2022}. Concurrently, architectures like M3GNet~\cite{chen2022m3gnet} showed the power of scaling, combining graph propagation and explicit three-body terms with broad training on the Materials Project~\cite{riebesell2025matbench}. Recently, the boundary between polynomial expansions and graph networks has blurred. MACE bridged this gap by combining ACE-inspired symmetric contractions with higher-order equivariant messages~\cite{batatia2022}. Graph ACE formalized the relationship between cluster correlations and semilocal graphs~\cite{bochkarev2024}. Finally, as these architectures mature, the focus has shifted toward robust physical execution, where models like eSEN utilize energy-derived forces, smooth cutoff envelopes, and continuous equivariant processing to demonstrate that static error metrics alone cannot guarantee molecular-dynamics stability or accurately capture physical observables~\cite{fu2025esen}.

% The atomic cluster expansion (ACE) gives a systematic description of local many-body geometry. It expands a neighbor density in radial functions and spherical harmonics, then couples angular momenta to form scalar and tensor correlations~\cite{drautz2019}. Equivariant neural networks use the same representation theory while learning nonlinear maps between irreducible tensor channels~\cite{geiger2022}. NequIP introduced these features into an interatomic graph network and showed the data efficiency of equivariant message passing~\cite{batzner2022}. M3GNet combined graph propagation, explicit
% three-body terms, and broad Materials Project~\cite{riebesell2025matbench} training for materials
% relaxation and dynamics~\cite{chen2022m3gnet}. MACE combined ACE-inspired symmetric contractions with higher-order equivariant messages~\cite{batatia2022}. Graph ACE then made the relation between cluster correlations and semilocal graphs explicit, and GRACE extended this view to broad materials data sets \cite{bochkarev2024,lysogorskiy2026}. eSEN combined energy-derived forces, smooth cutoff envelopes, and continuous equivariant tensor processing, and showed why static errors alone are not sufficient tests of molecular-dynamics stability or physical observables~\cite{fu2025esen}.

Attention mechanisms offer a powerful alternative to capture complex dependencies within the data. Originally introduced as a learned alignment strategy for sequence models~\cite{bahdanau2015}, this approach was subsequently formalized by the Transformer architecture, which integrated multi-head attention and feed-forward layers into stacked repeating blocks~\cite{vaswani2017}. \draftthirty{Set Transformer subsequently formulated multihead attention as a permutation-invariant operation on unordered sets, including attention pooling from a small set of learned queries~\cite{lee2019set}.} For atomistic systems, attention must also preserve permutation symmetry and three-dimensional transformation laws. DPA-1 uses gated attention in a local invariant potential~\cite{zhang2024dpa1}. The SE(3)-Transformer combines invariant attention weights with equivariant value messages~\cite{fuchs2020se3}, while TorchMD-NET couples distance-dependent attention to scalar and vector atomic features~\cite{tholke2022torchmd}. Equiformer embeds attention in an equivariant graph network with high-degree tensor features~\cite{liao2026}. SO3krates couples invariant atomic features to sparse spherical-harmonic variables while updating both representations across neighboring atoms~\cite{frank2022so3krates}. Attention and spatial propagation are separate choices. The Point Edge Transformer applies attention to edge tokens and restores rotational symmetry by a separate symmetrization step~\cite{pozdnyakov2023}. EScAIP uses optimized self-attention over scalar neighbor representations \cite{qu2024}. For pushing the boundaries of computational efficiency, these recent models along with Orb-v3~\cite{rhodes2025} have demonstrated that non-equivariant architectures can still accurately capture complex and higher-order physical properties. On the other hand, allegro follows a strictly local route in which equivariant ordered-pair features are refined without atom-centered message passing, and it has enabled large molecular-dynamics simulations~\cite{musaelian2023}. These models show that message passing, attention, the explicit many-body structure, and communication between atomic states can be varied independently.

TRACE adopts a fixed-environment factorization. ACE density correlations first summarize one cutoff environment as an equivariant center state. This state supplies the query in local cross-attention. The keys and equivariant values are built from the original directed-edge features and do not receive updated states from neighboring atoms. Additional attention blocks can therefore refine the nonlinear response within one environment without enlarging the spatial support of its atomic energy. ACE-correlated center queries fixed tensorial edge features through cutoff-preserving attention, and one invariant energy generates both forces and stress. We test this construction on three different chemical systems: polymorph stability and collective transformation in cesium lead iodide (CsPbI$_3$), partial pair structure in liquid water, and the activation free energy of an intramolecular reaction. Together they test structural relaxation, variable-cell dynamics, finite-temperature sampling, and bond rearrangement. 

%They are not matched comparisons with other MLIPs. Such comparisons require common data, partitions, parameter counts, optimization budgets, numerical precision, and hardware.

\section{Methods}

\begin{figure*}[t]
\centering
\includegraphics[width=0.98\textwidth]{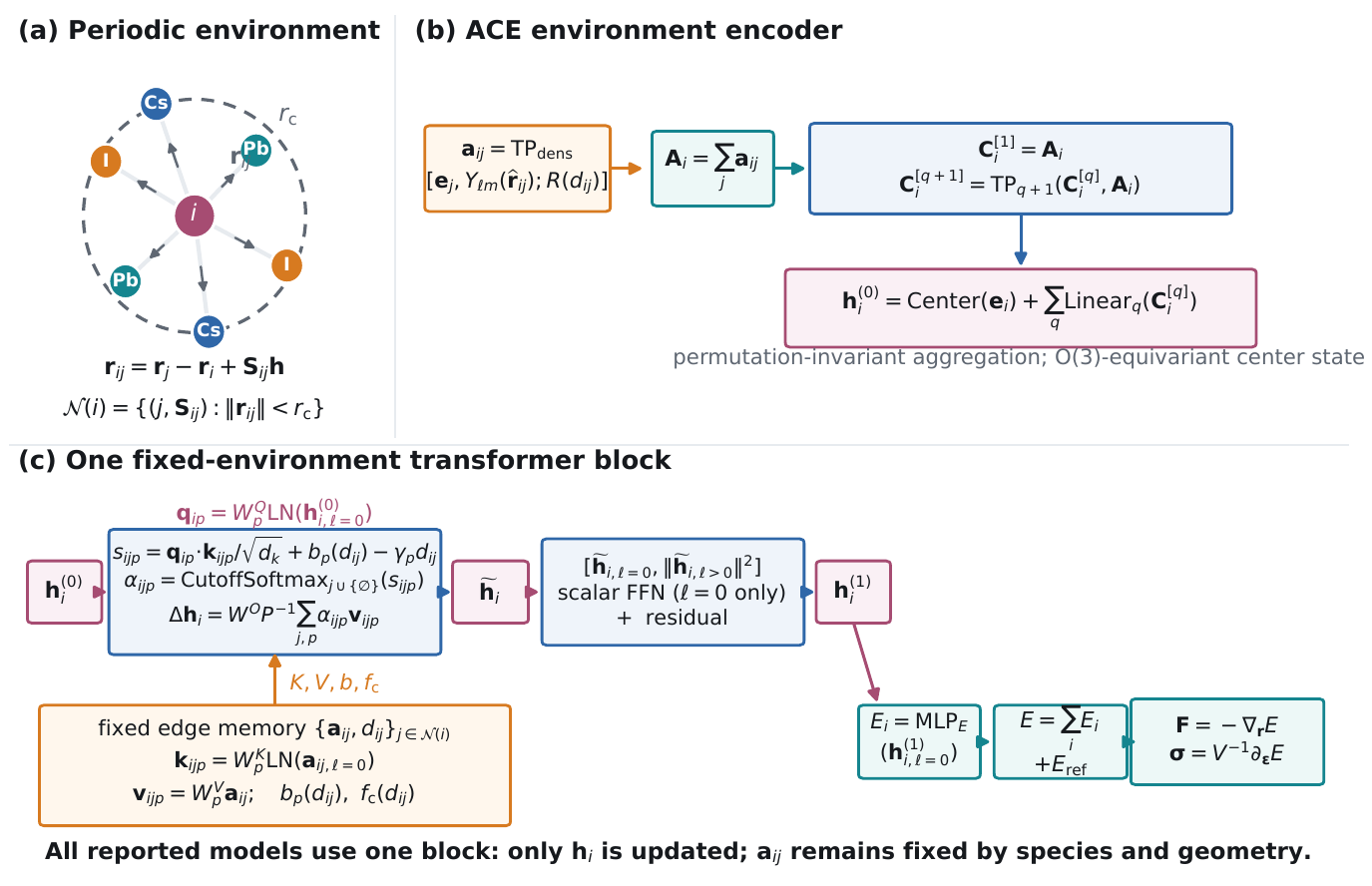}
\caption{\drafttwentyeight{TRACE architecture: (a) For receiver $i$, the vector $\vect r_{ij}=\vect r_j-\vect r_i+\vect S_{ij}\vect h$ points to a periodic image of sender $j$ inside the cutoff. (b) The sender species, radial basis, and spherical harmonics form a directed O(3) edge tensor $\vect a_{ij}$. Summation over incoming edges gives the neighbor density $\vect A_i$; learned Clebsch--Gordan products of that density generate the ACE correlations and initial center state. (c) One multihead cross-attention block constructs queries from even scalar center channels, keys from scalar edge channels, and values from the complete edge tensors. The cutoff-weighted softmax contains a null channel. An equivariant attention residual is followed by a feed-forward residual that updates only scalar channels and is conditioned on squared norms of the nonscalar tensors. The edge memory is unchanged. A scalar atomic readout and $E_{\mathrm{ref}}$ give the total energy; its derivatives give forces and stress.}}
\label{fig:trace-architecture}
\end{figure*}

\drafttwentyeight{Figure~\ref{fig:trace-architecture} follows the implemented TRACE architecture. For each receiver $i$, an image-resolved neighbor list defines vectors from $i$ to the selected periodic image of each sender atom $j$. The sender species and these distances and directions form O(3) edge tensors, whose receiver-wise sum and recursive Clebsch--Gordan products initialize the center state. In the single block used in every reported calculation, even scalar center channels provide the queries, scalar edge channels provide the keys, and the complete edge tensors provide the values. Only the center state is updated; the edge tensors remain functions of the input species and geometry. An invariant atomic readout and a composition-dependent reference give the total energy, from which forces and stress are obtained by differentiation.}

\subsection{Local energy decomposition and symmetry requirements}

TRACE writes the potential energy as a sum of atomic contributions and a composition-dependent reference,
\begin{equation}
 E(\mathcal R,\vect h)=\sum_{i=1}^{N}E_i+E_{\mathrm{ref}}(\{Z_i\}).
 \label{eq:energy-decomp}
\end{equation}
Here $Z_i$ is the species of atom $i$, and $E_i$ depends only
on its finite local environment. The reference may be a fitted sum of
per-species energies or a constant mean energy per atom. It is independent of coordinates and homogeneous strain and therefore contributes neither forces nor stress.

The total energy is invariant under translation, rotation, inversion, and permutation of atoms of the same species. Internal features need not be invariant; they may transform equivariantly. A feature with angular degree $\ell$ and parity $p=\pm1$ obeys
\begin{equation}
 \vect x^{(\ell,p)}\mapsto D^{(\ell,p)}(g)\vect x^{(\ell,p)},\qquad g\in O(3),
 \label{eq:irrep-transform}
\end{equation}
where $D^{(\ell,p)}(g)$ is an irreducible representation of
$O(3)$. For a proper rotation, it is the degree-$\ell$ Wigner representation; inversion adds the parity factor $p$. The energy readout uses only invariant scalars and invariant contractions. Its coordinate derivative therefore transforms as a vector under rotations and reflections. TRACE uses e3nn tensor products for this algebra~\cite{geiger2022}. An O(3)-invariant scalar assigns the same energy to isolated enantiomers related by reflection, as required for a parity-conserving potential without a chiral external field. Parity-breaking physics would require additional inputs and a different symmetry assumption.

To limit cost, TRACE retains only natural-parity irreducible
representations, $(\ell,p)=(\ell,(-1)^\ell)$ for
$0\leq\ell\leq\ell_{\max}$, and projects Clebsch--Gordan products back into this set. It therefore omits unnatural-parity intermediates such as even vectors and odd scalars. This is a compact truncation of the O(3) tensor space, not a complete set of all O(3)-allowed tensor features.

\subsection{Periodic local environments}

A periodic neighbor list must identify both an atom and its cell image. Each directed edge $j\rightarrow i$ therefore carries an integer shift $\vect S_{ij}\in\mathbb Z^3$. With row-vector coordinates and cell matrix
$\vect h$, TRACE uses the displacement
\begin{align}
 \vect r_{ij} &= \vect r_j-\vect r_i+\vect S_{ij}\vect h,\\
 d_{ij} &= \lVert\vect r_{ij}\rVert_2,\qquad
 \widehat{\vect r}_{ij}=\vect r_{ij}/d_{ij}.
 \label{eq:periodic-edge}
\end{align}
The corresponding neighbor set is
\begin{equation}
 \calN(i)=\{(j,\vect S_{ij}):d_{ij}<r_{\mathrm c}\}.
 \label{eq:neighbor-set}
\end{equation}
Every sum written as $\sum_{j\in\calN(i)}$ runs over these image-resolved edges. If several images of one atom lie within the cutoff, each is included separately. The same directed-edge convention and shifted displacements are used for training, inference, and LAMMPS deployment, including all spherical harmonics, radial functions, force derivatives, and strain derivatives.

\subsection{Smooth radial basis and cutoff regularity}

An edge must disappear smoothly when it reaches the cutoff; otherwise, neighbor-list changes introduce discontinuities in forces or higher derivatives. TRACE uses the compact quintic envelope
\begin{equation}
 f_{\mathrm c}(d)=
 \begin{cases}
 1-10x^3+15x^4-6x^5, & 0\le x<1,\\
 0, & x\ge 1,
 \end{cases}
 \qquad x=d/r_{\mathrm c}.
 \label{eq:cutoff}
\end{equation}
It satisfies
\begin{equation}
 f_{\mathrm c}(r_{\mathrm c})=f_{\mathrm c}'(r_{\mathrm c})
 =f_{\mathrm c}''(r_{\mathrm c})=0.
 \label{eq:cutoff-smoothness}
\end{equation}
Thus the value and its first two radial derivatives vanish at the boundary. Edge contributions remain continuous through second derivatives as neighbors enter or leave the list. TRACE expands distances in the cutoff-weighted basis
\begin{equation}
 B_n(d)=f_{\mathrm c}(d)\sqrt{\frac{2}{r_{\mathrm c}}}
 \frac{\sin(\omega_n d/r_{\mathrm c})}{d},\qquad n=1,\ldots,N_r,
 \label{eq:bessel}
\end{equation}
where $\omega_n=n\pi$ initially and may optionally be optimized. The implementation uses the analytic limit at $d\rightarrow0$. It also provides the Gaussian basis
\begin{equation}
 B_n^{\mathrm G}(d)=f_{\mathrm c}(d)
 \exp\!\left[-\frac{1}{2}\left(\frac{d-\mu_n}{s_n}\right)^2\right],
 \label{eq:gaussian}
\end{equation}
with fixed or trainable centers $\mu_n$ and widths $s_n$. All calculations reported here use fixed-frequency Bessel functions. We do not evaluate the Gaussian or trainable-frequency options.

\subsection{Chemical-angular ACE density}

Each directed edge is represented by chemical, radial, and
angular information. A learned embedding maps species $Z_i$ to even scalar
channels, $\vect e_i=\mathrm{Emb}(Z_i)$. Real spherical harmonics
$Y_{\ell m}(\widehat{\vect r}_{ij})$, in component normalization, describe the edge direction and transform with degree $\ell$ and parity $(-1)^\ell$. A bias-free radial network maps 
$\vect B(d_{ij})=(B_1,\ldots,B_{N_r})$ to the weights of an equivariant tensor product,
\begin{equation}
 \vect a_{ij}=
 \mathrm{TP}_{\mathrm{dens}}\!\left(
   \vect e_j,
   \{Y_{\ell m}(\widehat{\vect r}_{ij})\}_{\ell=0}^{\ell_{\max}};
   \mathrm{MLP}_{r}[\vect B(d_{ij})]
 \right).
 \label{eq:edge-density}
\end{equation}
Here $\mathrm{TP}_{\mathrm{dens}}$ constructs the edge channels; the semicolon separates its tensor inputs from its distance-dependent weights. In components,
\begin{equation}
 a^{(\ell,(-1)^\ell)}_{ij,cm}
 =Y_{\ell m}(\widehat{\vect r}_{ij})
 \sum_q R^{(\ell)}_{cq}[\vect B(d_{ij})]e_{j,q}.
 \label{eq:edge-density-components}
\end{equation}
The learned radial function $R^{(\ell)}_{cq}$ resolves
distance, while $e_{j,q}$ carries the identity of neighbor $j$. Because the radial input includes the cutoff and the radial network has no bias,
$\vect a_{ij}$ vanishes at $r_{\mathrm c}$. TRACE couples two equivariant
features with the Clebsch--Gordan product
\begin{equation}
 [\vect x\otimes_{\mathrm{CG}}\vect y]^{(\ell,p)}_{cm}
 =\sum_{\substack{\ell_1,p_1,c_1,m_1\\
                   \ell_2,p_2,c_2,m_2}}
 W^{\ell_1\ell_2\ell}_{c_1c_2c}
 C^{\ell m}_{\ell_1m_1,\ell_2m_2}
 x^{(\ell_1,p_1)}_{c_1m_1}y^{(\ell_2,p_2)}_{c_2m_2},
 \qquad p=p_1p_2.
 \label{eq:cg}
\end{equation}
The Clebsch--Gordan coefficients
$C^{\ell m}_{\ell_1m_1,\ell_2m_2}$ are fixed by symmetry, while $W$ mixes
learned channels. Allowed couplings obey
$|\ell_1-\ell_2|\le \ell\le \ell_1+\ell_2$ and $p=p_1p_2$.

The local neighbor density at center $i$ is the permutation-invariant sum
\begin{equation}
 \vect A_i=\sum_{j\in\calN(i)}\vect a_{ij}.
 \label{eq:density}
\end{equation}
\draftthirtytwo{Because the radial input is proportional to
$f_{\mathrm c}$ and the smooth bias-free radial network maps zero to zero,
$\vect a_{ij}=\mathcal O(f_{\mathrm c})$ as
$d_{ij}\rightarrow r_{\mathrm c}^{-}$. For the quintic cutoff,
$f_{\mathrm c}=\mathcal O[(r_{\mathrm c}-d_{ij})^3]$; every edge
contribution and its first two radial derivatives therefore vanish at the
cutoff.}

\subsection{Learned ACE correlations}

Equation~\eqref{eq:density} contains one factor of the
neighbor density. Repeated tensor products of this density generate many-body
correlations. TRACE constructs them recursively,
\begin{align}
 \vect C_i^{[1]} &= \vect A_i,\\
 \vect C_i^{[q+1]} &=
 \mathrm{TP}_{q+1}\!\left(\vect C_i^{[q]},\vect A_i\right),
 \qquad 1\le q<q_{\max}.
 \label{eq:ace-recursion}
\end{align}
where $q$ is the polynomial degree in the density. The
implementation names $q_{\max}+1$ \nolinkurl{correlation_order}. Counting the center, degree $q$ has nominal maximum body order $q+1$. Because products of a summed density include repeated neighbor indices, each
$\vect C_i^{[q]}$ also contains lower-body terms and is not a pure
$(q+1)$-body contribution. TRACE therefore uses a learned, compressed,
truncated ACE density-correlation basis. It does not enumerate a complete
linear ACE or symmetry-adapted $U$-matrix basis. Increasing $\ell_{\max}$, the radial resolution, the channel counts, or $q_{\max}$ increases model capacity, but the finite learned projections do not form a nested complete sequence and do not guarantee monotonic convergence to the complete ACE limit. Permutation symmetry follows because all correlations are formed from the neighbor sum
$\vect A_i$. The initial center state is
\begin{equation}
 \vect h_i^{(0)}=
 \mathrm{Center}(\vect e_i)
 +\sum_{q=1}^{q_{\max}}\mathrm{Linear}_{q}
 \!\left(\vect C_i^{[q]}\right).
 \label{eq:center-state}
\end{equation}
The center embedding enters the scalar channels, so two atoms
with similar neighbor densities remain distinguishable when their central
species differ. The quantity $q_{\max}+1$ describes only the nominal maximum
body order of the polynomial correlations before attention. The complete TRACE
energy has no finite polynomial body order because layer normalization,
exponential attention normalization, SiLU activations, and the nonlinear
readout act within the same local environment.

\subsection{Fixed-environment cross-attention}

The ACE correlations provide one state for each center.
TRACE updates this state with a transformer-style block containing multihead
scaled dot-product cross-attention, a residual connection, and a scalar
feed-forward sublayer. The operation is cross-attention from the center to its
directed edges, not self-attention among neighbor tokens and not graph
attention between updated atomic states. The center supplies the query, while
fixed edge tensors supply the keys and equivariant values. The implementation
supports $L$ blocks; every model reported here uses one. For block $t$ and head
$p$, the query and key are
\begin{align}
 \vect q^{(t)}_{ip} &= W^{Q,t}_p\,\mathrm{LN}\!\left(\vect h^{(t)}_{i,\ell=0}\right),\\
 \vect k^{(t)}_{ijp} &= W^{K,t}_p\,\mathrm{LN}\!\left(\vect a_{ij,\ell=0}\right).
 \label{eq:qk}
\end{align}
where $p=1,\ldots,H$ labels a head, $d_k$ is its query and
key dimension, and $\mathrm{LN}$ normalizes scalar channels. Both vectors are
invariant scalars with respect to spatial transformations, so their dot product
defines the invariant logit
\begin{equation}
 s^{(t)}_{ijp}=
 \frac{(\vect q^{(t)}_{ip})^{\mathsf T}\vect k^{(t)}_{ijp}}{\sqrt{d_k}}
 +b^{(t)}_p(\vect B(d_{ij}))
 -\mathrm{softplus}(\lambda^{(t)}_p)d_{ij},
 \qquad \widetilde s^{(t)}_{ijp}=s^{(t)}_{ijp}/T_{\mathrm{att}}(\tau).
 \label{eq:logit}
\end{equation}
Here $b^{(t)}_p$ is a learned radial bias,
$\mathrm{softplus}(\lambda^{(t)}_p)$ is a nonnegative distance coefficient, and
$T_{\mathrm{att}}(\tau)>0$ is a dimensionless attention temperature at training
epoch $\tau$. The subscript distinguishes it from thermodynamic temperature.
Its training schedule is stored with the run, and
$T_{\mathrm{att}}=1$ for all reported inference calculations. TRACE includes a
unit null contribution when normalizing each neighbor set,
\begin{equation}
 \alpha^{(t)}_{ijp}=
 \frac{f_{\mathrm c}(d_{ij})\exp(\widetilde s^{(t)}_{ijp})}
 {1+\sum_{k\in\calN(i)}f_{\mathrm c}(d_{ik})
 \exp(\widetilde s^{(t)}_{ikp})}.
 \label{eq:cutoff-attention}
\end{equation}
so the cutoff factor is not canceled by normalization when only one edge is present. Consequently,
$\alpha^{(t)}_{ijp}\rightarrow0$ as $d_{ij}\rightarrow r_{\mathrm c}$, and an
edge leaves the attention sum smoothly.

The value for each head is an equivariant linear map of the
fixed edge tensor,
\begin{equation}
 \vect v^{(t)}_{ijp}=\mathrm{Linear}^{V,t}_{p}(\vect a_{ij}),
 \label{eq:value}
\end{equation}
and the head-averaged update is added to the center through a
residual connection,
\begin{equation}
 \vect u^{(t)}_i=
 \mathrm{Linear}^{O,t}\!\left[
 \frac{1}{H}\sum_{p=1}^{H}\sum_{j\in\calN(i)}
 \alpha^{(t)}_{ijp}\vect v^{(t)}_{ijp}\right],
 \qquad
\widetilde{\vect h}_i^{(t)}=
 \vect h_i^{(t)}+\vect\gamma^{(t)}_{\mathrm{attn}}\odot\vect u^{(t)}_i.
 \label{eq:attention-update}
\end{equation}
\draftthirtytwo{The smooth bias-free radial network gives
$\vect v_{ijp}^{(t)}=\mathcal O(f_{\mathrm c})$ near the cutoff, while
$\alpha_{ijp}^{(t)}$ contains a second, explicit cutoff factor. Each edge contribution to the update is therefore
$\mathcal O(f_{\mathrm c}^{\,2})$, rather than an exact algebraic product of two cutoff factors. Within each irrep copy, the effective residual scale
$\vect\gamma^{(t)}_{\mathrm{attn}}$ is shared over all magnetic (quantum number index) components $m$. It is an invariant scalar multiplier and preserves the tensor transformation law; $\odot$ denotes multiplication by these tied scales. Training applies dropout to the attention weights before aggregation, whereas validation and inference do not.}

The attention residual is followed by a feed-forward update of the scalar channels. For each nonscalar irrep copy in $\widetilde{\vect h}_i^{(t)}$, TRACE first forms the invariant squared norm
\begin{equation}
 n^{(t)}_{ic\ell}=\sum_{m=-\ell}^{\ell}
 \left|\widetilde h^{(t,\ell)}_{icm}\right|^2,
 \qquad \ell>0.
 \label{eq:norm}
\end{equation}
The even scalar channels and these norms are then combined as
\begin{align}
 \vect\chi_i^{(t)}&=\mathrm{LN}\!\left[
 \widetilde{\vect h}^{(t)}_{i,\ell=0},
 \{n^{(t)}_{ic\ell}\}_{\ell>0}\right],\\
 \Delta\vect s_i^{(t)}&=\mathrm{MLP}^{(t)}_{\mathrm{FFN}}
 \!\left(\vect\chi_i^{(t)}\right),\\
 \vect h^{(t+1)}_{i,\ell=0}&=
 \widetilde{\vect h}^{(t)}_{i,\ell=0}
 +\vect\gamma^{(t)}_{\mathrm{FFN}}\odot\Delta\vect s_i^{(t)},\\
 \vect h^{(t+1)}_{i,\ell>0}&=
 \widetilde{\vect h}^{(t)}_{i,\ell>0}.
 \label{eq:scalar-ffn}
\end{align}
Tensor norms can therefore influence later scalar queries and
the energy, but no componentwise nonlinearity is applied to nonscalar irreps.
\draftthirtytwo{The implemented scalar MLP has the sequence
linear--SiLU--dropout--linear--dropout. The two dropout operations use the
same probability as attention dropout and are disabled during validation and
inference.}
The normalization over the local neighbor set and the scalar feed-forward
network are nonlinear. The complete model consequently has no finite effective
body order even though its pre-attention density correlations do.

The query depends on the current state of center $i$, whereas
the keys and values depend only on fixed descriptors $\vect a_{ij}$. No updated
state $\vect h_j^{(t)}$ is sent from atom $j$ to atom $i$. The spatial support of
each atomic energy is therefore exactly $r_{\mathrm c}$ for any number of
blocks. A force on atom $i$ differentiates every site energy whose environment
contains $i$ and can consequently depend on pairs of atoms separated by up to
about $2r_{\mathrm c}$. Domain decomposition still needs only one
$r_{\mathrm c}$ ghost halo for each owned center, followed by reverse
communication of forces on ghost atoms.

\subsection{Fixed-environment dependency}

The dependency can be compared directly with message passing.
A generic graph update has the form
\begin{equation}
 \vect h_i^{(t+1)}
 =\Phi\!\left(\vect h_i^{(t)},
 \sum_{j\in\calN(i)}
 \Psi(\vect h_i^{(t)},\vect h_j^{(t)},\vect r_{ij})\right),
 \label{eq:message-passing-form}
\end{equation}
in which the current state of neighbor $j$ enters the update
of center $i$. TRACE instead uses
\begin{equation}
 \vect h_i^{(t+1)}
 =\Phi_{\mathrm{loc}}\!\left(
 \vect h_i^{(t)},
 \{\vect a_{ij}\}_{j\in\calN(i)}\right),
 \qquad
 \vect a_{ij}=\vect a(Z_j,\vect r_{ij}),
 \label{eq:trace-local-form}
\end{equation}
where every $\vect a_{ij}$ is built once from the input
species and geometry. The edge set is a fixed geometric memory, while only the center state changes between blocks. Equation~\eqref{eq:trace-local-form} is therefore local cross-attention rather than an exchange of learned sender states.

\subsection{Invariant readout, forces, and stress}

After $L$ attention blocks, let $\bar{\vect h}_i=\vect h_i^{(L)}$ be the final center state. Only its even scalar channels enter the atomic-energy readout,
\begin{equation}
 E_i=\mathrm{MLP}_{E}\!\left(\bar{\vect h}_{i,\ell=0}\right),
 \qquad E=\sum_iE_i+E_{\mathrm{ref}}(\{Z_i\}).
 \label{eq:energy}
\end{equation}
Forces are derivatives of this scalar energy, not separate network outputs,
\begin{equation}
 \vect F_i=-\left.\frac{\partial E}{\partial\vect r_i}\right|_{\vect h}.
 \label{eq:force}
\end{equation}
The resulting force field is conservative up to numerical
precision and neighbor-list tolerances. For a periodic structure with positive volume, stress is the derivative of the same energy with respect to symmetric homogeneous strain.
Define
\begin{equation}
 \vect\epsilon(\vect\eta)=
 \begin{pmatrix}
 \eta_1&\eta_4&\eta_5\\
 \eta_4&\eta_2&\eta_6\\
 \eta_5&\eta_6&\eta_3
 \end{pmatrix},\qquad
 \vect\Lambda(\vect\eta)=\vect I+\vect\epsilon(\vect\eta).
 \label{eq:strain}
\end{equation}
Here $\vect\Lambda$ is the deformation map, not an atomic force. Applying it to both the cell and Cartesian coordinates keeps fractional coordinates fixed,
\begin{equation}
 \vect h'=\vect h\vect\Lambda,\qquad
 \vect r_i'=\vect r_i\vect\Lambda.
 \label{eq:strain-deformation}
\end{equation}
At zero strain, $V=|\det\vect h|$. In the ASE convention used
for the training labels and calculator, the Cauchy stress is
\begin{align}
 \sigma_{aa}&=\left.\frac{1}{V}\frac{\partial E}{\partial\eta_a}
 \right|_{\vect\eta=0},
 &&a=1,2,3,\\
 \sigma_{xy}&=\left.\frac{1}{2V}\frac{\partial E}{\partial\eta_4}
 \right|_{\vect\eta=0},\quad
 \sigma_{xz}=\left.\frac{1}{2V}\frac{\partial E}{\partial\eta_5}
 \right|_{\vect\eta=0},\quad
 \sigma_{yz}=\left.\frac{1}{2V}\frac{\partial E}{\partial\eta_6}
 \right|_{\vect\eta=0}.
 \label{eq:stress}
\end{align}
The factor $1/2$ appears because one shear parameter changes
two symmetric off-diagonal entries. In the ASE sign convention,
$P=-\operatorname{tr}(\vect\sigma)/3$, and the corresponding LAMMPS virial is
$\vect W=-V\vect\sigma$. Finite differences of all six strain components and
the deployed virial path are tested against these equations.

\subsection{Training objective and data partitioning}

Energies constrain the value of the learned surface, while forces and stress constrain its derivatives. All three can therefore enter one training objective. For batch $\mathcal B$, let $I_s^\sigma=1$ when structure
$s$ has a stress label and zero otherwise. TRACE minimizes
\begin{align}
 \calL_E&=\frac{1}{|\mathcal B|}\sum_{s\in\mathcal B}
 \left[\frac{E_s-E_s^{\mathrm{ref}}}{N_s}\right]^2,\\
 \calL_F&=\frac{1}{|\mathcal B|}\sum_{s\in\mathcal B}
 \frac{1}{3N_s}\sum_{i,a}
 \left(F_{sia}-F_{sia}^{\mathrm{ref}}\right)^2,\\
 \calL_\sigma&=\frac{1}{6|\mathcal B|}
 \sum_{s\in\mathcal B}I_s^\sigma\sum_{v=1}^{6}
 \left(\sigma_{sv}-\sigma_{sv}^{\mathrm{ref}}\right)^2,\\
 \calL&=w_E\calL_E+w_F\calL_F+w_\sigma(t)\calL_\sigma
 +w_{\mathrm{Sob}}\calL_{\mathrm{Sob}}.
 \label{eq:loss}
\end{align}
The superscript ``ref'' denotes the reference label for a
given data set. The energy error is divided by the number of atoms before it is squared. Force errors are averaged over the $3N_s$ Cartesian components of each
structure. Thus every structure has equal weight, regardless of size; this is not the same as averaging all atomic components over a mixed-size data set.
Structures without stress labels contribute zero to $\calL_\sigma$, but its denominator remains the full batch size. Its effective contribution therefore
depends on the labeled fraction of the batch. The optional local-linearization
term is
\begin{equation}
 \calL_{\mathrm{Sob}}=
 \frac{1}{|\mathcal B|}\sum_{s\in\mathcal B}\left[
 E_s(\vect r_s+\vect\delta_s)-E_s(\vect r_s)
 +\sum_{i=1}^{N_s}\vect F_{si}\cdot\vect\delta_{si}
 \right]^2,\qquad
 \delta_{sia}\sim\mathcal N(0,\sigma_\delta^2),
 \label{eq:sobolev}
\end{equation}
where the force in the linear term is detached from this
auxiliary graph. The stress weight may be increased gradually during the first epochs to avoid large second-derivative updates before the energy surface has learned its basic local shape.

Reported errors are defined independently of the loss
weights. With $\Delta E_s=E_s-E_s^{\mathrm{ref}}$, the energy RMSE is
\begin{equation}
 \mathrm{RMSE}_E=
 \left[\frac{\sum_s N_s(\Delta E_s/N_s)^2}{\sum_sN_s}\right]^{1/2}.
 \label{eq:energy-metric}
\end{equation}
The force RMSE is the square root of the structure-averaged component mean-square error,
\begin{equation}
 \mathrm{RMSE}_F=
 \left[\frac{1}{|\mathcal S|}\sum_{s\in\mathcal S}\frac{1}{3N_s}
 \sum_{i,a}(\Delta F_{sia})^2\right]^{1/2}.
 \label{eq:force-metric}
\end{equation}
The stress RMSE includes the six Voigt components of
structures that carry stress labels,
\begin{equation}
 \mathrm{RMSE}_\sigma=
 \left[\frac{1}{6|\mathcal S_\sigma|}
 \sum_{s\in\mathcal S_\sigma}\sum_{v=1}^{6}
 (\Delta\sigma_{sv})^2\right]^{1/2}.
 \label{eq:stress-metric}
\end{equation}
We report energy RMSE in meV per atom, force RMSE in eV~\AA$^{-1}$, and stress RMSE in eV~\AA$^{-3}$.

Adjacent molecular-dynamics frames are correlated. TRACE
therefore uses blocked splitting by default: contiguous blocks are assigned to validation, and a gap around each block is omitted to reduce temporal leakage. This is more demanding than a random frame split, but independent trajectories and phases are still needed to test transferability. 

\subsection{Optimization, checkpointing, and scaling}

TRACE supports AdamW/AMSGrad and the Muon matrix optimizer
\cite{jordan2024muon}, with AdamW for auxiliary parameter groups. The Muon update and auxiliary path follow the open-source Nequix implementation~\cite{koker2025nequix}; TRACE uses the parameter grouping stated below. Because Muon was used for the CsPbI$_3$ model presented in the following sections, we give the implemented update explicitly. For a matrix gradient $\vect G_t$ and momentum $\vect M_t$,
\begin{align}
 \vect M_t&=\beta\vect M_{t-1}+(1-\beta)\vect G_t,\\
 \vect H_t&=(1-\beta)\vect G_t+\beta\vect M_t,
 \label{eq:muon-momentum}
\end{align}
with $\vect H_t=\vect M_t$ instead when Nesterov momentum is
disabled. If $\vect H_t$ has shape $m\times n$, it is transposed temporarily
when $m>n$. The normalized matrix is
\begin{equation}
 \vect X_0=\frac{\vect H_t}{\lVert\vect H_t\rVert_F+10^{-7}},
 \label{eq:muon-normalization}
\end{equation}
Five quintic Newton--Schulz steps then approximate its matrix
zero-power map,
\begin{align}
 \vect A_k&=\vect X_k\vect X_k^{\mathsf T},\\
 \vect X_{k+1}&=a\vect X_k+
 (b\vect A_k+c\vect A_k^2)\vect X_k,
 \label{eq:muon-ns}
\end{align}
with $(a,b,c)=(3.4445,-4.7750,2.0315)$. After restoring the
original orientation, the result is multiplied by
$\sqrt{\max(1,m/n)}$ to give $\vect U_t$. With learning rate $\eta$ and
decoupled weight decay $\lambda$,
\begin{equation}
 \vect\Theta_{t+1}=(1-\eta\lambda)\vect\Theta_t-\eta\vect U_t.
 \label{eq:muon-update}
\end{equation}
For the CsPbI$_3$ model, Muon updates only the query, key, and hidden scalar feed-forward matrices. AdamW updates embeddings, radial and tensor-product parameters, normalization parameters, biases, residual scales, and the energy readout. Both groups use $\eta=10^{-3}$ and
$\lambda=10^{-5}$. Muon uses $\beta=0.95$; AdamW uses coefficients
$(0.9,0.95)$ and $\epsilon=10^{-10}$. Newton--Schulz operations use float32 in
the reported CPU run. The run manifest records the learning-rate schedule,
gradient clipping, attention-temperature schedule, stress-weight ramp,
optimizer state, and data split.

Deployment retains only the scalar-energy graph. ASE and
LAMMPS obtain forces and virials by differentiating that graph. For LAMMPS, a
fixed type map is stored when the PyTorch checkpoint is exported to a
TorchScript/LibTorch artifact and loaded by the native pair style. Strict
locality means that domain decomposition needs the usual cutoff ghost atoms,
but no repeated exchange of hidden states between ranks.

For mean neighbor count $\overline n$, the evaluation cost has the form
\begin{equation}
 \mathcal O\!\left[N\overline n
 \left(C_{\mathrm{edge}}+L H C_{\mathrm{attn}}\right)\right]
 +\mathcal O\!\left(N C_{\mathrm{CG}}+N C_{\mathrm{read}}\right),
 \label{eq:scaling}
\end{equation}
where $L$ and $H$ are the numbers of blocks and heads.
$C_{\mathrm{edge}}$, $C_{\mathrm{attn}}$, $C_{\mathrm{CG}}$, and
$C_{\mathrm{read}}$ denote the costs of an edge tensor, one attention head,
the local correlations, and the readout. For bounded $\overline n$, the total
work is linear in $N$. More blocks increase local work but do not enlarge the
cutoff environment or domain halo. This asymptotic result is not a hardware
benchmark. Actual accelerator and multi-GPU performance also depends on
neighbor lists, tensor kernels, derivatives, memory movement, and
communication.

\section{Applications}
Calculating errors on novel test structures do not by themselves establish that a machine learning potential reproduces the physics and chemistry needed in simulations to calculate experimentally measurable observables in physical and chemical systems. We therefore test TRACE on three different problems: relative energies of polymorphs, finite-temperature stability, and hard-core phase transformation in CsPbI$_3$; partial radial distribution functions in liquid water; and the activation free energy of a chemical reaction. Each chemical system is trained separately with the similar local TRACE architecture and observables are compared against experiments. 

\subsection{Crystallization: polymorph energetics, phase diagram, and rare-events}

CsPbI$_3$ is a fully inorganic halide perovskite with immense potential for solar energy. Its ``black'' perovskite phase possesses a band gap of $\simeq$1.7 to 1.8~eV, which is nearly ideal for the top absorbing layer in a perovskite--silicon tandem solar cell---a design capable of pushing theoretical solar to power conversion efficiencies beyond 40\%~\cite{bremner2016bandgaps,futscher2016tandem}. However, a major practical challenge is that this photo-active black phase is metastable. It naturally competes with a structurally stable, but yellow and photo-inactive, non-perovskite $\delta$ phase. At the atomic level, the inactive $\delta$ phase consists of edge-sharing PbI$_6$ octahedra, whereas the active black phases (whether orthorhombic, tetragonal, or cubic) are defined by corner-sharing octahedra with varying tilt angles.

During both the synthesis and daily operation of these solar cells, the material frequently transitions between these polymorphs. Simulating this transition is uniquely difficult because it involves sweeping changes: the Pb--I connectivity breaks and reforms, the local environment around the cesium atoms shifts entirely, and the overall volume and shape of the simulation cell changes continuously. Therefore, to be successful, a potential must accomplish two things: it must correctly predict the relative thermodynamic stability of each polymorph, and it must guarantee smooth, continuous forces and stresses throughout massive structural disruptions.

To train a TRACE potential capable of handling this complexity, we utilized our previously established r$^2$SCAN+rVV10 density functional theory (DFT) calculations~\cite{ahlawat2024crystallization}. This dataset provides high-fidelity energies, forces, and stresses across 979 snapshots of periodic 96-atom CsPbI$_3$ supercells of all polymorphs. We rigorously divided this data into 863 training structures and 100 validation structures. To ensure the model learns the underlying physics rather than simply memorizing closely correlated molecular dynamics snapshots, we intentionally discarded 16 boundary frames to cleanly separate the training and validation sets. The model was trained entirely on this fixed dataset in a single pass, without relying on any active learning steps. The complete model architecture and training parameters are summarized in Table~\ref{tab:cspbi3-architecture}.

\begin{table}[p]
\centering
\small
\renewcommand{\arraystretch}{1.12}
\begin{tabular}{p{0.44\linewidth}p{0.43\linewidth}}
\toprule
Quantity & Value \\
\midrule
Cutoff / radial basis & 6.0~\AA{} / 12 Bessel functions \\
Maximum angular degree & $\ell_{\max}=2$ \\
Node irreps & $64\times0e+32\times1o+16\times2e$ \\
Correlation irreps & $16\times0e+8\times1o+4\times2e$ \\
{\raggedright Pre-attention density degree / nominal maximum body order\par}
  & 3 / 4 \\
Radial network & $12\rightarrow32\rightarrow$ TP weights \\
Attention & 1 local block, 2 heads \\
Scalar FFN & invariant scalars and $\ell>0$ squared norms \\
Readout & $64\rightarrow64\rightarrow1$ \\
Trainable parameters & 132,005 \\
Optimizer & Muon for hidden matrices; AdamW auxiliary groups \\
Learning rate / weight decay & $10^{-3}$ / $10^{-5}$ \\
Batch size / epochs & 8 / 100 \\
Loss weights $(w_E,w_F,w_\sigma)$ & $(1,10,10^3\rightarrow10^5)$ \\
Stress-weight ramp & 20 epochs \\
Local-linearization $(w_{\mathrm{Sob}},\sigma_\delta)$ & $(10^{-3},0.02~\text{\AA})$ \\
Attention dropout / layer scale & 0.03 / 0.01 \\
Direct force/stress heads & none \\
Long-range electrostatics & not included \\
\bottomrule
\end{tabular}
\caption{Architecture and training parameters for the CsPbI$_3$ model. The nominal body order counts the center and applies only to the polynomial density correlations before attention. Repeated neighbor indices also produce lower-body terms. Attention and the nonlinear readout remove a finite body-order interpretation of the complete model.}
\label{tab:cspbi3-architecture}
\end{table}
\FloatBarrier

\begin{figure}[htbp]
\centering
\includegraphics[width=0.82\linewidth]{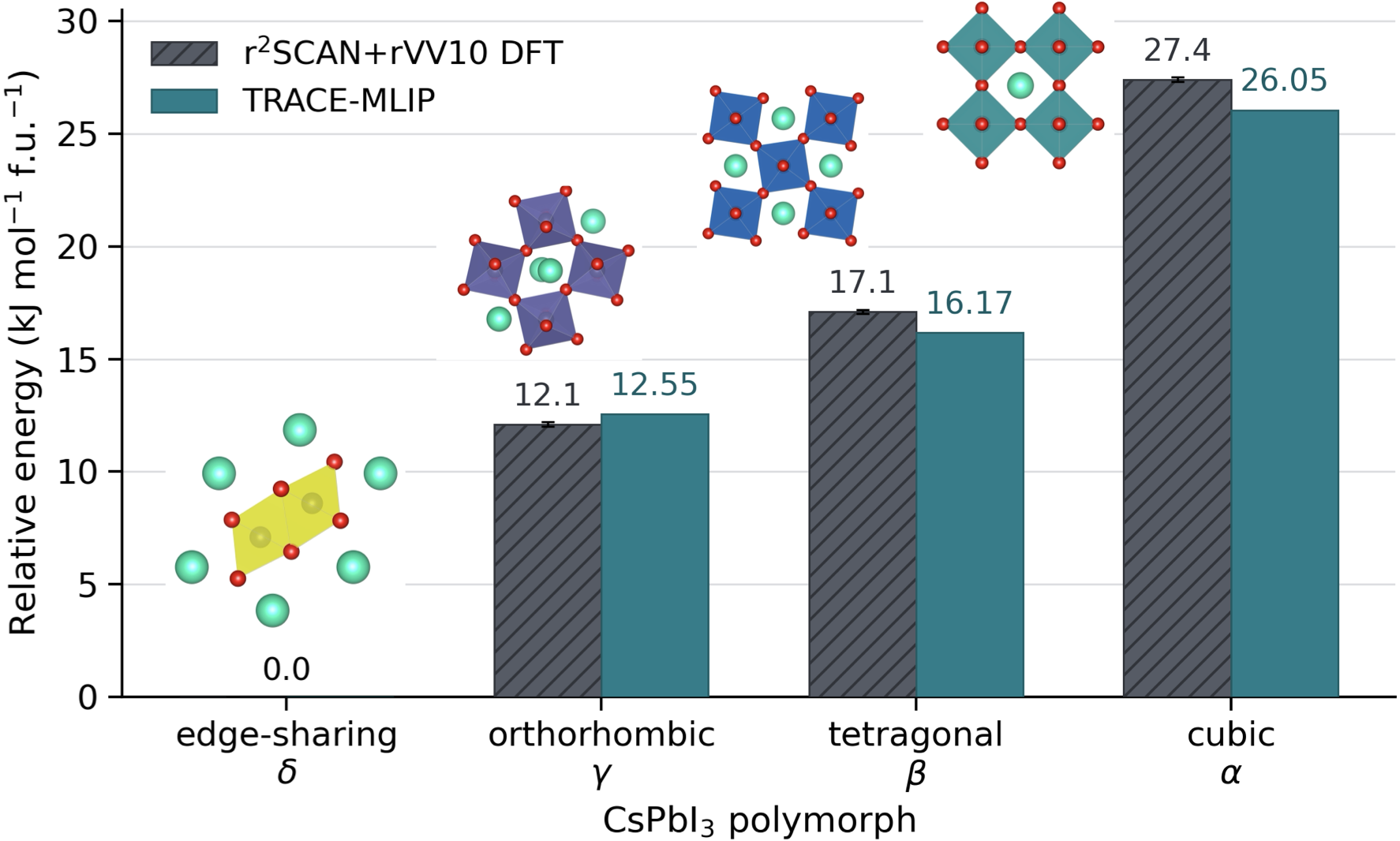}
\caption{\drafttwentysix{Relative energies of independently relaxed CsPbI$_3$ polymorphs from r$^2$SCAN+rVV10 DFT and TRACE-MLIP, in kJ~mol$^{-1}$ per formula unit relative to the edge-sharing $\delta$ phase.}}
\label{fig:cspbi3-relative-energies}
\end{figure}

After training, we first test the zero-temperature relative energies after relaxing both the atomic positions and the cell of each polymorph. The checkpoint with the lowest validation loss, obtained at epoch 76, was used to relax 20-atom cells of the edge-sharing $\delta$, orthorhombic $\gamma$, tetragonal $\beta$, and cubic $\alpha$ phases. Every relaxation reached a maximum force below 0.01~eV~\AA$^{-1}$. Relative to the $\delta$ phase, TRACE-MLIP gives energies of 0, 12.55, 16.17, and 26.05~kJ~mol$^{-1}$ per formula unit for $\delta$, $\gamma$, $\beta$, and $\alpha$, respectively. The corresponding r$^2$SCAN+rVV10 values are approximately 0, 12.1, 17.1, and 27.4~kJ~mol$^{-1}$ per formula unit [Fig.~\ref{fig:cspbi3-relative-energies}]. TRACE-MLIP preserves the phase ordering, and its deviations from DFT are 0.45, $-0.93$, and$-1.35$~kJ~mol$^{-1}$ per formula unit for $\gamma$, $\beta$, and $\alpha$. The DFT and TRACE structures were relaxed independently; the comparison therefore tests relaxed phase energies rather than single-point energies at a common geometry.

\FloatBarrier

% The DFT values were read from phase-energy bars reporting $E_\delta-E_p$ and were sign-reversed to the convention shown.

\begin{figure}[htbp]
\centering
\includegraphics[width=\linewidth]{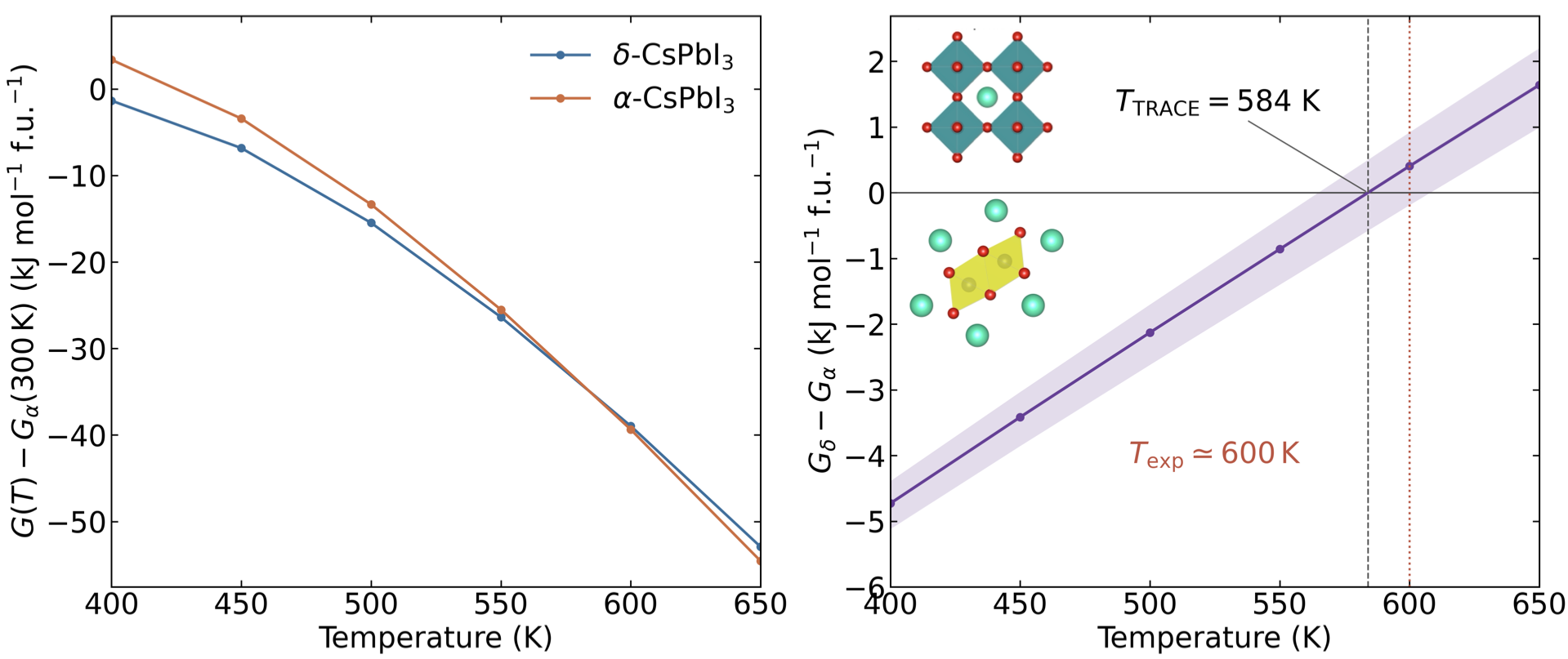}
\caption{\drafttwentysix{Finite-temperature stability of edge-sharing
$\delta$-CsPbI$_3$ and cubic $\alpha$-CsPbI$_3$ at 1.01325~bar. Left: Gibbs
free energies relative to $G_\alpha(300~\mathrm{K})$, obtained from a
Frenkel--Ladd reference at 450~K and Gibbs--Helmholtz integration. Right:
$\Delta G_{\delta\alpha}=G_\delta-G_\alpha$; negative values favor $\delta$s
and positive values favor $\alpha$. The shaded band is the nominal 95\% block-and-replica bootstrap interval. Vertical lines mark the calculated crossing at $\sim$580~K and the experimental completion of the cubic transformation near 600~K. The calculation uses 240-atom cells and classical nuclei.}}
\label{fig:cspbi3-phase-diagram}
\end{figure}

However, relative energies alone do not determine which phase is stable at finite temperatures and pressures. The stable phase minimizes the Gibbs free energy, $G=H-TS$, where entropy can change the zero-temperature phase diagrams. Therefore, we calculated the absolute free energies of 480-atom edge-sharing $\delta$ and cubic $\alpha$ cells at $T_0=450$~K and $p=1.01325$~bar. For each phase $s\in\{\delta,\alpha\}$, the TRACE energy $U_s(\mathbf R)$ was connected to the energy of an Einstein crystal through the Frenkel--Ladd path
\cite{frenkel1984}
\begin{equation}
\drafttwentysixeq{
\begin{aligned}
 U_{\lambda,s}(\mathbf R)
 &= (1-\lambda)U_s(\mathbf R)
    +\lambda U_{\mathrm{Ein},s}(\mathbf R),\\
 U_{\mathrm{Ein},s}(\mathbf R)
 &= \frac{1}{2}\sum_i \kappa_{Z_i,s}
    \left|\mathbf r_i-\mathbf r^0_{i,s}\right|^2,
 \qquad 0\leq\lambda\leq1 .
\end{aligned}}
\label{eq:cspbi3-frenkel-ladd-path}
\end{equation}
\drafttwentysix{where $\mathbf R=\{\mathbf r_i\}$ contains all atomic
positions, $\mathbf r^0_{i,s}$ is the reference site of atom $i$, and
$\kappa_{Z_i,s}$ is the spring constant for its chemical species. The reference volume $V_s$ is the mean $NpT$ volume at $(T_0,p)$, and the spring constants are obtained from the mean-square displacements in each phase. The path has the TRACE solid at $\lambda=0$ and the Einstein crystal at $\lambda=1$. After removing center-of-mass motion, the Helmholtz free energy of phase $s$ is}
\begin{equation}
\drafttwentysixeq{
 F_s(T_0,V_s)=F_{\mathrm{Ein},s}(T_0,V_s)
 +\int_0^1
 \left\langle U_s-U_{\mathrm{Ein},s}\right\rangle_{\lambda,s}
 \,d\lambda .}
\label{eq:cspbi3-frenkel-ladd-free-energy}
\end{equation}
For each phase, five independent pairs of 30-ps forward and
reverse paths were combined with the symmetric work estimator, which reduces the leading error from finite switching rates \cite{freitas2016}. From $G_s(T_0,p)=F_s(T_0,V_s)+pV_s$, the free energy at other temperatures follows from the Gibbs--Helmholtz relation
\begin{equation}
\drafttwentysixeq{
 \frac{G_s(T,p)}{T}
 =\frac{G_s(T_0,p)}{T_0}
 -\int_{T_0}^{T}\frac{H_s(T',p)}{T'^2}\,dT',
 \qquad
 H_s=\left\langle U_s+K+pV\right\rangle_{NpT}.}
\label{eq:cspbi3-gibbs-helmholtz}
\end{equation}
The enthalpy was sampled every 50~K from 300 to 650~K.
Figure~\ref{fig:cspbi3-phase-diagram} shows the resulting change in
stability. At 400~K, $\Delta G_{\delta\alpha}=G_\delta-G_\alpha=-4.87$~kJ~mol$^{-1}$ per formula unit, so the $\delta$ phase is stable. At 650~K, $\Delta G_{\delta\alpha}=+1.92$~kJ~mol$^{-1}$ per formula unit, and the $\alpha$ phase is stable. Linear interpolation gives $\Delta G_{\delta\alpha}=0$ at $\simeq$580~K. High-temperature x-ray diffraction experiments find $\delta$ / cubic coexistence from 563 to 602~K and a fully cubic phase at 602~K \cite{trots2008}; the calculated crossing lies within the experimental observations. The nominal 95\% block-and-replica bootstrap interval is 553--599~K. This interval measures statistical uncertainty from the sampled trajectories, but not the remaining forward--reverse hysteresis of 0.7--4.7~meV per atom or the error from the single 480-atom cell. The value $\sim$580~K is therefore a preliminary classical estimate. Within these limits, TRACE reproduces both the zero-temperature relative energies and finite-temperature phase diagrams. 

Moving beyond stable equilibrium phases, we next test whether the TRACE potential can describe a highly complex transition relevant to industrial applications: the $\delta$-to-perovskite transformation in CsPbI$_3$. This process requires edge-sharing PbI$_6$ octahedra to break apart and form a corner-sharing network. At the same time, the local environments of the Pb, I, and Cs atoms must change alongside the dimensions of the simulation cell. Earlier simulations have shown that this transformation produces complicated intermediate structures, including mixed-connectivity layers and stacking faults~\cite{ahlawat2024crystallization} and later observed in various experiments~\cite{chen2025ripening, chen2025heterophase, song2025intragrain, dong2026intermediate}. We therefore designed a test to see if our fixed potential could drive this entire collective reorganization of this multi-species system.

To follow the transformation, we employ biased simulations using multi-species structure-factor reaction coordinate, adapted from our previous work on CsPbI$_3$ crystallization \cite{ahlawat2024crystallization}. Let $\mathcal A_\alpha$ denote the atoms of species $\alpha\in\{\mathrm{I},\mathrm{Pb},\mathrm{Cs}\}$ and let $r_{ij}$ be the minimum-image distance from a Pb center $i$ to atom $j$. The species-resolved local response is:
\begin{align}
 s_i^{(\alpha)}
 &=1+\sum_{\substack{j\in\mathcal A_\alpha,\;j\ne i\\
                     r_{ij}<r_{\mathrm c}^{(\alpha)}}}
 j_0(q_\alpha r_{ij})
 j_0\!\left(\frac{\pi r_{ij}}{r_{\mathrm c}^{(\alpha)}}\right),
 \qquad
 j_0(x)=\frac{\sin x}{x},\quad j_0(0)=1,
 \label{eq:cspbi3-local-structure-factor}\\
 \chi_\alpha(u)
 &=\frac{(u/s_\alpha^\star)^6}
         {1+(u/s_\alpha^\star)^6},
 \qquad
 S_{\mathrm p}
 =\sum_{i\in\mathcal A_{\mathrm{Pb}}}
 \prod_{\alpha\in\{\mathrm{I},\mathrm{Pb},\mathrm{Cs}\}}
 \chi_\alpha\!\left(s_i^{(\alpha)}\right).
 \label{eq:cspbi3-collective-variable}
\end{align}
Equation~\eqref{eq:cspbi3-local-structure-factor} uses a spherical Bessel function to probe the specific perovskite length scale, while a second function smoothly brings each pair contribution to zero at its radial cutoff. The parameters are:
\begin{equation}
 \begin{split}
 (q_{\mathrm I},q_{\mathrm{Pb}},q_{\mathrm{Cs}})
 &=(21.99,23.78,14.36)\ \mathrm{nm}^{-1},\\
 r_{\mathrm c}^{(\alpha)}&=1.2\ \mathrm{nm},\\
 (s_{\mathrm I}^\star,s_{\mathrm{Pb}}^\star,s_{\mathrm{Cs}}^\star)
 &=(1.5,1.3,1.5).
 \end{split}
 \label{eq:cspbi3-collective-variable-parameters}
\end{equation}
The switch function, $\chi_\alpha$, maps each the local response toward zero or one. When multiplied together, their product is large only when the I, Pb, and Cs environments around a single Pb center. Summing this across all Pb centers gives $S_{\mathrm p}$, an extensive measure of a multi-species crystalline structure. Because it uses only distances between identical species, $S_{\mathrm p}$ is independent of cell translation and rotation. 

\begin{figure}[htbp]
\centering
\includegraphics[width=0.78\linewidth]{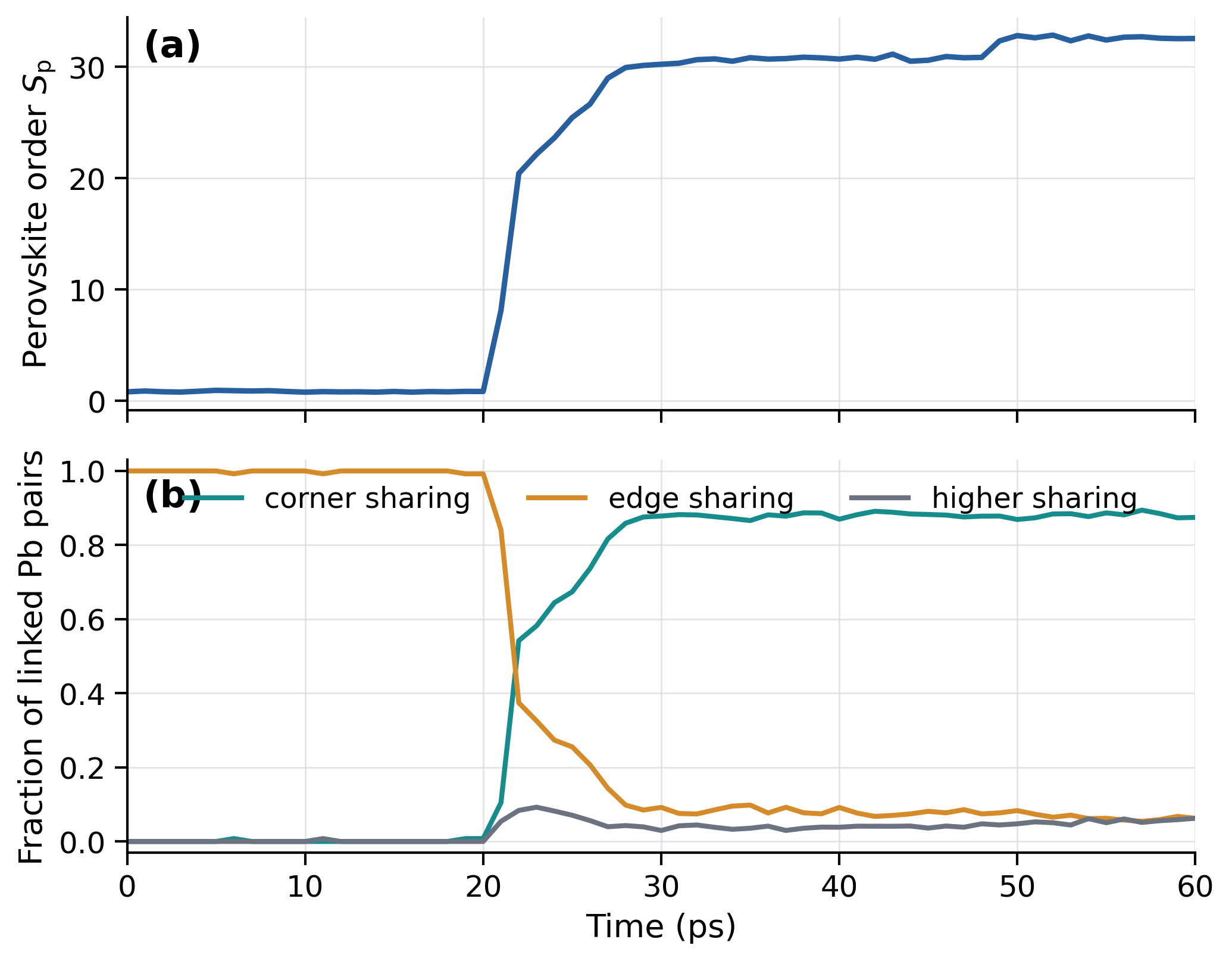}
\caption{Collective phase transformation during the
enhanced-sampling transition in CsPbI$_3$. (a) Pb-centered perovskite order
$S_{\mathrm p}$. (b) Fractions of linked Pb pairs sharing one iodide
(corner sharing), two iodides (edge sharing), or three or more iodides
(higher sharing). Connectivity is defined with a 4.2-\AA{} Pb--I cutoff.}
\label{fig:cspbi3-rare-event}
\end{figure}

To test the model, we prepared a 640-atom system (128 formula units) in the edge-sharing $\delta$ phase. After 2~ps of thermalization, we ran the simulation at 400~K and 1~bar with a 2-fs time step, allowing all cell lengths and angles to fluctuate for reaching equilibrium. We then applied an expanded-ensemble bias potential using overlapping umbrellas along $S_{\mathrm p}$ for sampling between the $\delta$ and perovskite phases~\cite{plumed2014,invernizzi2020}. Crucially, this bias only changes statistical weights and sample free energy surface; where the underlying TRACE energy surface remained unchanged. Throughout the trajectory, every energy, force, and cell derivative was evaluated using the same frozen checkpoint, without any active learning or additional electronic-structure calculations.

The simulation demonstrates a clear structural transition [Fig.~\ref{fig:cspbi3-rare-event}(a)]. For the first 20~ps, the system remains in the non-perovskite basin with $S_{\mathrm p}$ near 0.8. The order parameter then increases to 8.13 at 21~ps and reaches 29.92 by 28~ps, eventually stabilizing between 30 and 33 for the remainder of the 60-ps run. To verify that this change reflects a true physical rewiring of the PbI$_6$ network, we tracked how iodides were shared between linked Pb pairs. Initially, all 256 linked pairs share an edge between all PbI$_6$ octahedra. By 60~ps, the network has completely transformed: 293 pairs are corner-sharing, leaving only 21 edge-sharing pairs and 21 pairs that share three or more iodides. This means corner-sharing pairs grow to account for 87.5\% of the linked network. The fact that the mixed corner-edge-sharing structure coexist during the simulations indicates that the cell reorganizes heterogeneously, rather than shifting all at once [Fig.~\ref{fig:cspbi3-rare-event}(b)]. This simultaneous increase in both $S_{\mathrm p}$ and corner-sharing connectivity proves that the fixed potential remains stable while the PbI$_6$ network, all three sublattices, and the simulation cell completely reorganize. The model successfully navigates into a new structural basin without needing to be trained on the intermediate configurations. 

We note that this biased trajectory demonstrates a collective rare event; it does not provide a physical time sequence or a specific microscopic mechanism. Extracting a quantitative free-energy barrier would require converged sampling, independent replicas, and electronic-structure validation outside the training distribution.

\FloatBarrier

\subsection{Liquids: water structure from a reduced CCSD(T) reference set}

Water is fundamental to chemistry and biology. For machine learning interatomic potentials, it represents a unique challenge. Unlike the rigid lattice of a crystalline solid, liquid water relies on a constantly fluctuating hydrogen-bond network. To succeed here, a model cannot just hold a structure together; it must remain stable while accurately capturing dynamical, finite-temperature correlations. To see how well TRACE handles this disorder, we trained it on a deliberately restricted subset of the widely used MB-pol coupled-cluster dataset. Our goal is to establish a classical baseline for the TRACE architecture using limited data, not to compete with the extensive sampling or nuclear-quantum corrections found in dedicated water models like NEP-MB-pol~\cite{xu2025}.

To create this stress test, we stripped down the available 1250-frame NEP-MB-pol dataset to 417 configurations. We split this data into 359 training structures and 50 validation structures, carefully discarding eight boundary frames to prevent the model from memorizing adjacent, highly correlated molecular dynamics snapshots. The TRACE model---configured with a 6-\AA{} cutoff, $\ell_{\max}=2$, 64 hidden scalar channels, and two attention heads---was trained on CPUs. We selected the checkpoint with the lowest validation loss (epoch 77), which yielded errors of 7.22~meV~atom$^{-1}$ for energy, 0.110~eV~\AA$^{-1}$ for forces, and $4.93\times10^{-4}$~eV~\AA$^{-3}$ for stress. 

\begin{figure}[htbp]
\centering
\includegraphics[width=0.68\linewidth]{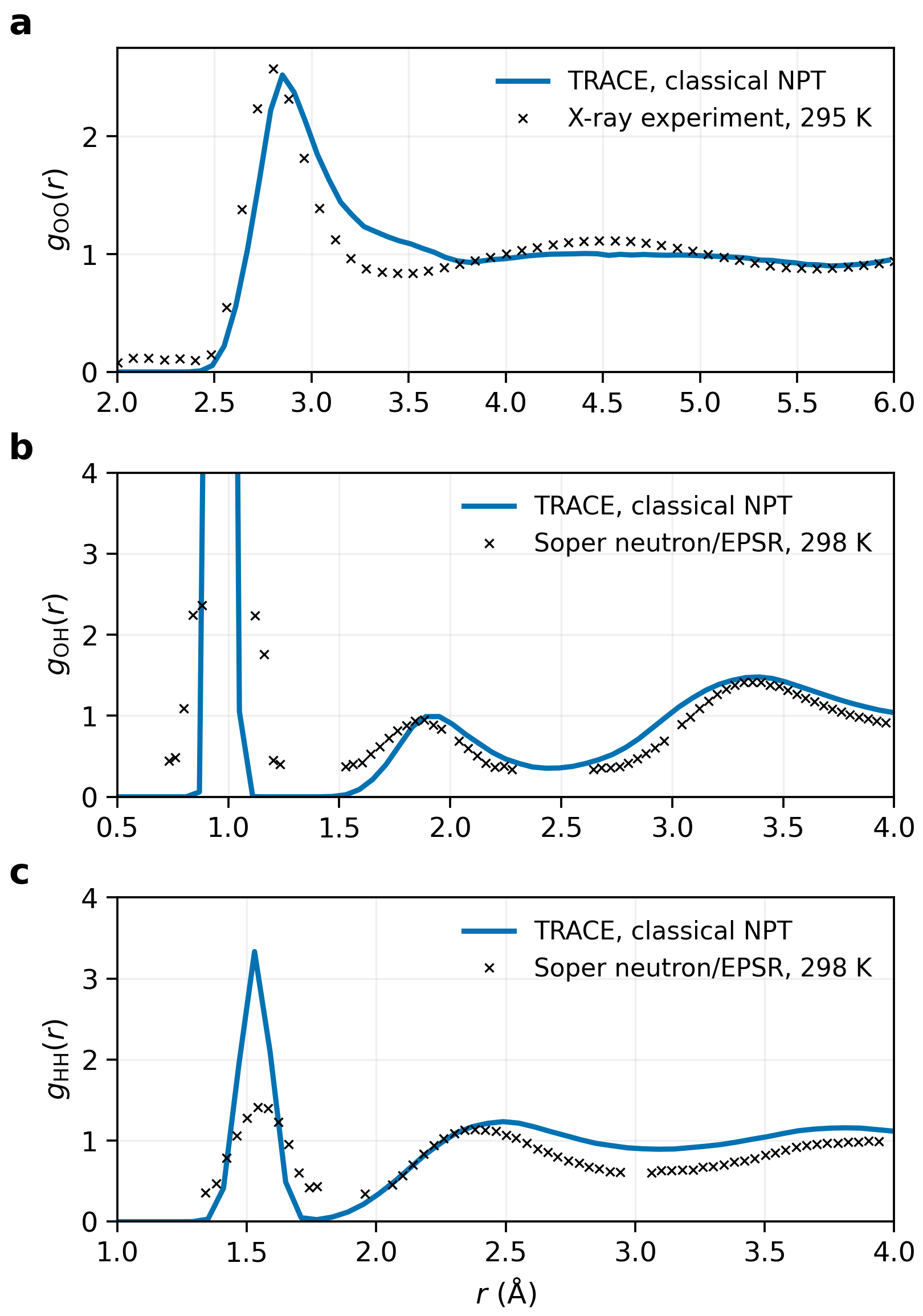}
\caption{Partial radial distribution functions from a
144-molecule classical TRACE trajectory at 300~K and 1~bar: (a) O--O,
(b) O--H, and (c) H--H. Solid curves show the final cumulative averages.
Crosses in (a) are every eighth point from the 295-K x-ray O--O data of
Skinner \emph{et al.}~\cite{skinner2013}. Crosses in (b) and (c) are Soper's 298-K neutron-diffraction/EPSR functions~\cite{soper2000}, read from Fig.~1 of Ref.~\cite{xu2025}. Panel (b) is limited to $g_{\mathrm{OH}}=4$ so that the intermolecular structure remains visible; the intramolecular O--H peak lies above this range.}
\label{fig:water-oo-rdf}
\end{figure}

We deployed this potential in LAMMPS to simulate a periodic box of 144 water molecules. We performed a classical isothermal-isobaric ($NPT$) simulations at 300~K and 1~bar for 135.2~ps trajectory, using a 0.5-fs time step. Figure~\ref{fig:water-oo-rdf} compares our resulting partial radial distribution functions against established experimental data. For a model trained on such a limited dataset, this agreement is highly encouraging. TRACE places the first oxygen--oxygen (O--O) peak at 2.85~\AA{} with a height of 2.52. This closely tracks the ambient-water x-ray data from Skinner \emph{et al.}~\cite{skinner2013}, which locates the peak at 2.80~\AA{} with a height of 2.58. Across the broader intermolecular range ($2.2\leq r\leq 6.0$~\AA), the pointwise RMSE between our predicted O--O structure and the experimental curve is just 0.217.

While the first O--O peak aligns well with experiment, structural agreement alone does not guarantee a perfect thermodynamic model. Because this was a purely classical simulation, it inherently misses the nuclear quantum effects that naturally soften the structure of real water. Furthermore, this relatively short trajectory was primarily designed to capture local correlations, meaning the mean density and time-averaged pressure are not yet fully converged. A truly quantitative validation of the liquid equation of state requires independent configurations, longer production intervals, checks for finite-size effects, and matching quantum-thermodynamic conditions. Most importantly, matching an $NPT$ radial distribution function is only the first step; the pressure and virial must ultimately be verified independently under variable-cell dynamics to confirm the macroscopic stability of the liquid.

\subsection{Chemical reactions: activation free energy of methyl migration}

\begin{figure}[p]
\centering
\includegraphics[width=0.98\linewidth]{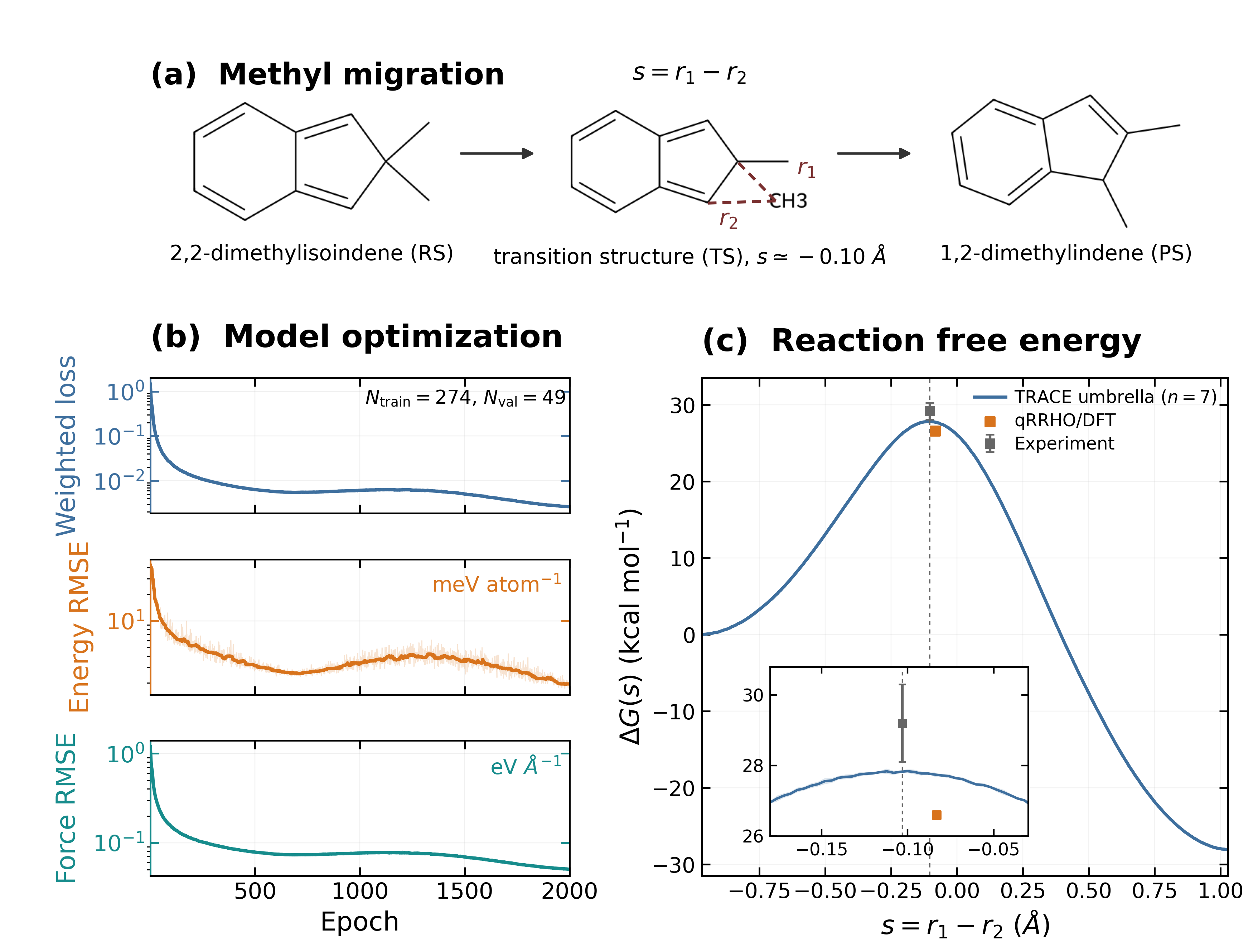}
\caption{Description of methyl migration from
2,2-dimethylisoindene to 1,2-dimethylindene. (a) Reaction scheme; dashed
contacts define the distances in Eq.~\eqref{eq:methyl-shift-coordinate}.
(b) Training errors for the fixed 274/49 split. Pale lines are epoch values; opaque lines are 31-epoch moving medians. (c) Umbrella-sampling profile at 365.6~K, shown as the mean and pointwise 95\% Student-$t$ interval over seven replicas. Squares mark the experimental barrier in pentane and the PBE0-D3BJ/def2-SVP qRRHO estimate. These are scalar barrier references, not profiles along $s$.}
\label{fig:trace-methyl-shift}
\end{figure}

While the solid and liquid examples validate structural stability without altering covalent connectivity, chemical reactions present a stricter test of breaking and formation a chemical bond. To determine whether this same local construction can accurately capture transition-state kinetics and free-energy profiles during such bond reorganization, we model a fundamentally important class of organic reactions: methyl migration. Specifically, we examine the intramolecular rearrangement of 2,2-dimethylisoindene to 1,2-dimethylindene [Fig.~\ref{fig:trace-methyl-shift}(a)], a process highly sensitive to kinetic barriers. One C--C bond weakens as another forms, so the trajectory samples geometries between the two stable connectivities. Experiments in pentane give an activation free energy of $29.2\pm1.1$~kcal~mol$^{-1}$ at 365.6~K \cite{dolbier1979,manning1981,vitartas2026}.

Vitartas \emph{et al.} released reference configurations at the PBE0-D3BJ/def2-SVP level~\cite{vitartas2026}. We joined 192 structures from their inherited-bias well-tempered-metadynamics active-learning calculation with 131 structures from downhill active learning. A fixed split assigns 274 structures to training and 49 to validation. We fitted TRACE once to this combined dataset, without another active-learning cycle or new electronic-structure labels. The model has 132,005 trainable parameters, a 5.0-\AA{} cutoff, $\ell_{\max}=2$, 12 radial functions, pre-attention density degree three (nominal maximum body order four), and one two-head fixed-environment attention block. Muon updates the hidden matrices and AdamW the remaining parameters. Full-precision CPU training used batches of eight for 2000 epochs. At the final checkpoint, the training energy RMSEs are 1.94~meV~atom$^{-1}$; the corresponding force RMSEs are 0.050 eV~\AA$^{-1}$, see [Fig.~\ref{fig:trace-methyl-shift}(b)].

Using this potential, we calculate the free-energy profile with umbrella sampling~\cite{torrie1977}. We define a reaction coordinate, $s$, using the distances from the migrating methyl carbon ($\vect r_{\mathrm m}$) to its original ($\vect r_{\mathrm o}$) and new ($\vect r_{\mathrm n}$) attachment sites:
\begingroup
\ifshowrevisions\color{draftsevencolor}\fi
\begin{equation}
 r_1=\left|\vect r_{\mathrm m}-\vect r_{\mathrm o}\right|,
 \qquad
 r_2=\left|\vect r_{\mathrm m}-\vect r_{\mathrm n}\right|,
 \qquad
 s=r_1-r_2.
 \label{eq:methyl-shift-coordinate}
\end{equation}
\endgroup
Here, negative and positive values of $s$ correspond to the reactant and product states, respectively. To fully sample the free energy along this coordinate, we apply a harmonic bias potential to each umbrella window $k$:
\begingroup
\ifshowrevisions\color{draftsevencolor}\fi
\begin{equation}
 U_k^{\mathrm{bias}}(s)
 =\frac{\kappa}{2}\left(s-s_k\right)^2,
 \qquad \kappa=20~\mathrm{eV}\,\text{\AA}^{-2}.
 \label{eq:methyl-shift-umbrella}
\end{equation}
\endgroup

\clearpage
To ensure we capture the transition state structure, we took 30 previously published window centers (from $-0.9685$ to $1.0283$~\AA) and added nine new midpoint windows, giving 39 windows in total. We simulated each gas-phase trajectory at 365.6~K for 40~ps with a 0.5-fs time step. Initial velocities were drawn independently from a Maxwell--Boltzmann distribution, and we discarded the first 10~ps of each run for equilibration. 

We then used the weighted histogram analysis method (WHAM)~\cite{kumar1992wham} to remove the sampling biases and reconstruct the unbiased probability density, $P(s)$. The free-energy profile, relative to the reactant minimum ($s_{\mathrm{RS}}$), is calculated as:
\begingroup
\ifshowrevisions\color{draftsevencolor}\fi
\begin{equation}
 \Delta G(s)=-k_{\mathrm B}T
 \ln\!\left[\frac{P(s)}{P(s_{\mathrm{RS}})}\right].
 \label{eq:methyl-shift-free-energy}
\end{equation}
\endgroup
While simulating an isolated molecule at a constant temperature technically yields a Helmholtz free-energy difference, we use the standard symbol $\Delta G$ to match conventional notations.

We perform seven independent runs of umbrella sampling, and the activation free energies are highly consistent among all of them, ranging from 27.860 to 27.967~kcal~mol$^{-1}$. The mean is $27.923 \pm 0.031$~kcal~mol$^{-1}$. The peak of the free-energy profile occurs between $s=-0.1186$ and $-0.0955$~\AA, with an average position of $-0.1048$~\AA{} that aligns well with the optimized transition state structure. We can be confident in these results because the underlying robust sampling: the smallest overlap between adjacent window histograms remains around 40\% (0.399--0.413) across all replicas. Furthermore, throughout all 273 trajectories, the molecular geometries remain stable, never violating our geometric limits ($r_1+r_2 < 3.20$~\AA{} or methyl C--H $> 1.35$~\AA). Ultimately, this means the smoothness of our free-energy profile comes from genuinely well-connected data and reproducible sampling, not from artificial, post hoc smoothing.

When we compare the TRACE barrier to existing benchmarks, the results are highly consistent. Our value sits 1.31~kcal~mol$^{-1}$ higher than the static PBE0-D3BJ/def2-SVP qRRHO estimate (26.6~kcal~mol$^{-1}$) and 1.29~kcal~mol$^{-1}$ lower than the experimental average. It is also very close to the $28.2 \pm 0.1$~kcal~mol$^{-1}$ result from the original potential~\cite{vitartas2026}. We must note that the experiment was performed in a solvent, and the qRRHO method relies on a stationary-point approximation, whereas our TRACE results are classical gas-phase simulation. Nevertheless, the tight agreement across our replicas proves that this model provides a highly reproducible activation energy for the chemical bond breaking and formation during methyl migration. To eventually extend this to a full solvent-phase rate constant, future work would need to define a standard state, verify convergence over longer trajectories, and account for dynamical recrossing beyond our one-dimensional coordinate.

% The seven activation free energies span 27.860--27.967~kcal~mol$^{-1}$. Their mean is $27.923\pm0.031$~kcal~mol$^{-1}$, where the uncertainty is a 95\% Student-$t$ interval over independent replicas. Profile maxima lie between $s=-0.1186$ and $-0.0955$~\AA, with mean $-0.1048$~\AA, close to the optimized transition structure. The smallest overlap between adjacent window histograms is 0.399--0.413 across the replicas. None of the 273 trajectories crosses the run-time geometry limits $r_1+r_2<3.20$~\AA{} or a methyl C--H distance above 1.35~\AA. The smooth profile is therefore supported by connected window histograms and consistent replicas rather than post hoc smoothing. 

% The TRACE barrier is 1.31~kcal~mol$^{-1}$ above the
% PBE0-D3BJ/def2-SVP qRRHO value of 26.6~kcal~mol$^{-1}$ and
% 1.29~kcal~mol$^{-1}$ below the experimental central value. It is also close to
% the $28.2\pm0.1$~kcal~mol$^{-1}$ umbrella result for the source ACE
% potential~\cite{vitartas2026}. These quantities are not equivalent: the
% experiment used pentane, qRRHO is a stationary-point approximation, and the
% TRACE result is a classical gas-phase profile. The agreement among replicas
% shows that the fixed model gives a reproducible activation free energy for this
% bond rearrangement. A solvent-phase rate constant would also
% require a defined standard state, convergence with trajectory length and
% replica count, and a treatment of dynamical recrossing beyond the
% one-dimensional coordinate.
\FloatBarrier

\section{Scope and limitations}

Three limits define the present scope. First, TRACE retains
natural-parity irreducible representations and projects every density-correlation degree into a finite learned channel space. It does not enumerate a complete linear ACE or $U$-matrix basis. The nominal four-body description applies only before attention; normalization and nonlinear scalar maps remove a finite body-order interpretation of the final energy. The effect of this truncation should be tested by varying parity content, angular and radial resolution, correlation degree, and channel count separately, and by measuring physical observables. Energy and harmonic errors alone do not establish accuracy for anharmonic thermomechanical response \cite{pota2024thermal}.

Second, the present energy has a finite cutoff and no explicit Coulomb, Ewald, charge-equilibration, dispersion-tail, or reciprocal-space term. For ionic CsPbI$_3$, this assumption must be tested against cutoff and cell size, long-wavelength distortions, dielectric environment, and charged defects. A long-range term should be introduced only if controlled tests reveal a systematic error that improved local data cannot remove.

Third, locality does not by itself prove practical speed or
transferability. For bounded density, TRACE has linear asymptotic work and one geometric halo, but this does not establish an accuracy or throughput advantage over another MLIP. The current TorchScript/LibTorch interface is not evidence of GPU-resident million-atom performance. Such a claim requires matched training, device-resident profiling, energy--force--virial agreement, memory
measurements, and strong- and weak-scaling tests on specified hardware. 

% Blocked validation also cannot replace independent trajectories and phases. For methyl migration, the stated uncertainty measures variation among seven replicas for one model and coordinate. It excludes errors from the reference method, training set, nuclear quantum effects, solvent, and reaction coordinate. Agreement with the solution-phase barrier is therefore not a gas-to-solution transfer test or a rate prediction.

\section{Discussion and conclusions}

TRACE is built around one dependency: an ACE-correlated
center state queries tensorial edge features that remain fixed by the input
species and geometry. ACE, O(3) equivariance, attention, and local energy
decomposition all have established precedents. TRACE combines them so that
nonlinear processing remains inside one cutoff environment and forces and
stress remain derivatives of a single invariant energy.

The applications test this construction through physical
observables rather than fit errors alone. For CsPbI$_3$, TRACE reproduces the
r$^2$SCAN+rVV10 ordering of four relaxed polymorphs, gives a preliminary classical phase crossing at $\simeq$580~K, and remains stable during conversion from an edge-sharing crystal to predominantly corner-sharing order. A separate water model places the first O--O maximum close to the diffraction reference. For methyl migration, seven independent umbrella calculations give
$27.92\pm0.03$~kcal~mol$^{-1}$ while sampling C--C bond rearrangement. With these results, we show that the same fixed-environment construction can be applied for wide range of crystalline, liquid, and reactive systems.

%With these results, our aim is not to establish superiority over other models or transferability beyond the sampled environments. 
% The next tests should be matched ablations and benchmarks.
% With descriptor, cutoff, parameter count, precision, and optimization budget held fixed, cross-attention should be compared with uniform weighting, no attention, and an ACE--MLP readout. Comparisons with established MLIPs should
% use common data partitions and report molecular-dynamics stability, equations of state, virials, throughput, memory, and multi-device scaling as well as energy and force errors. These measurements will determine when fixed-
% environment attention improves accuracy at useful cost and when information must propagate beyond one neighborhood.
\section*{Data and Software Availability}
The implementation, training configurations, and tests are
available at \href{https://github.com/paramvir3/Transformers-ACE}{\nolinkurl{github.com/paramvir3/Transformers-ACE}} and Zenodo \href{https://doi.org/10.5281/zenodo.21631673}{doi:10.5281/zenodo.21631673}

\section*{Acknowledgement}
The author acknowledges the use of OpenAI Codex in the development of this project.

% The equations in this article correspond to \nolinkurl{flashace/physics.py} and \nolinkurl{flashace/model.py}, including the fixed-environment descriptor, equivariant attention, invariant readout, and energy-derived force and stress paths. The methyl-migration reference structures and electronic-structure labels are available in the versioned archive associated with Ref.~\cite{vitartas2026}. The source repository contains the corresponding split manifest, training configuration, analysis, and figure scripts. A versioned public archive containing the fixed splits, reference configurations, checkpoints, trajectories, and plotting tables for all figures is in preparation. Until a persistent identifier is assigned, these materials are available from the author upon reasonable request.

% The references are complete inline; prevent RevTeX from also requesting a
% generated <jobname>Notes.bbl under latexmk/Overleaf.
\makeatletter
\let\auto@bib@innerbib\@empty
\makeatother

\clearpage
\appendix
\begin{draftthirtyoneblock}

\section{Fixed-environment tensorial cross-attention}
\label{app:attention}

This Appendix states the TRACE attention algorithm at the level of its tensor
dependencies and sparse implementation. The purpose is to distinguish three
ideas that are easily conflated: restricting atomic interactions to a finite
geometric neighborhood, applying attention to the tokens in that
neighborhood, and transmitting a learned hidden state from one atomic center
to another. TRACE uses the first two operations but not the third. Its block
is most precisely described as sparse local center-to-edge cross-attention
over fixed equivariant ACE density tokens.

\subsection{Center states and fixed directed-edge tokens}

Let \(N\) be the number of atoms and let \(\mathcal E\) be the image-resolved
directed neighbor list,
\begin{equation}
 \mathcal E=
 \left\{e=(j\!\rightarrow\! i,\vect S_{ij}):
 d_e=\left\lVert\vect r_j-\vect r_i+\vect S_{ij}\vect h\right\rVert
 <r_{\mathrm c}\right\}.
 \label{eq:app-edge-set}
\end{equation}
Its size is \(N_{\mathrm e}=|\mathcal E|\). For each edge \(e\), \(s(e)=j\)
denotes the sender and \(r(e)=i\) the receiver or central atom. Periodic images
are distinct entries when they lie inside the cutoff.

The ACE encoder first evaluates the cutoff-weighted radial vector
\begin{equation}
 \vect B(d_e)=f_{\mathrm c}(d_e)
 \bigl(\phi_1(d_e),\ldots,\phi_{N_r}(d_e)\bigr)
 \label{eq:app-radial-vector}
\end{equation}
and maps it through the bias-free radial network
\begin{equation}
 \vect w_e=\mathrm{MLP}_{r}\!\left[\vect B(d_e)\right].
 \label{eq:app-radial-weights}
\end{equation}
The resulting weights parameterize the directed equivariant token
\begin{equation}
 \vect a_e=\vect a_{ij}
 =\mathrm{TP}_{\mathrm{dens}}\!\left[
 \vect e_{Z_j},\,
 \{Y_{\ell m}(\widehat{\vect r}_e)\}_{\ell m};
 \vect w_e
 \right],
 \label{eq:app-edge-token}
\end{equation}
where \(\vect e_{Z_j}\) is the sender-species embedding,
\(\phi_n\) is a Bessel or Gaussian primitive, and the semicolon separates the
tensor inputs from the external tensor-product weights.
\begin{draftthirtytwoblock}
Because \(\vect B(r_{\mathrm c})=\vect 0\), the radial network is smooth and
has no additive bias, and its SiLU activation maps zero to zero, its complete
map \(F_r\) satisfies \(F_r(\vect 0)=\vect 0\). For finite network parameters
and radial primitives that are smooth at \(r_{\mathrm c}\),
\begin{equation}
 \vect w_e=F_r[\vect B(d_e)]
 =J_{F_r}(\vect 0)\vect B(d_e)
 +\mathcal O(\lVert\vect B(d_e)\rVert^2)
 =\mathcal O(f_{\mathrm c}(d_e)),
 \qquad
 \vect a_e=\mathcal O(f_{\mathrm c}(d_e)).
 \label{eq:app-cutoff-order}
\end{equation}
Thus the nonlinear radial network does not algebraically factor into
\(f_{\mathrm c}\) times a distance-only network; Eq.~\eqref{eq:app-cutoff-order}
is the precise asymptotic statement. Since the quintic envelope satisfies
\(f_{\mathrm c}(d)=\mathcal O[(r_{\mathrm c}-d)^3]\) from inside the cutoff,
\(\vect w_e\), \(\vect a_e\), and their first two radial derivatives vanish at
\(r_{\mathrm c}\).
\end{draftthirtytwoblock}
Receiver-wise summation gives the local neighbor density,
\begin{equation}
 \vect A_i=\sum_{e:r(e)=i}\vect a_e .
 \label{eq:app-density}
\end{equation}
The learned Clebsch--Gordan correlations of \(\vect A_i\), together with the
central-species embedding, form the initial center state
\(\vect h_i^{(0)}\) in Eq.~\eqref{eq:center-state}. The edge tokens
\(\{\vect a_e\}\) are then retained as a fixed input-derived memory during all
attention blocks in that energy evaluation. They are recomputed whenever the
atomic configuration or cell changes. They can depend on learned parameters,
species, distances, directions, and the periodic cell, but not on an
attention-updated state \(\vect h_j^{(t)}\).

For the illustrative \(\ell_{\max}=2\) construction with
\(\texttt{correlation\_channels}=16\), the edge representation is
\begin{equation}
 16\times0e\oplus8\times1o\oplus4\times2e,
 \qquad
 D_a=16+8(3)+4(5)=60.
 \label{eq:app-edge-width}
\end{equation}
With \(\texttt{hidden\_dim}=64\), the center representation is
\begin{equation}
 64\times0e\oplus32\times1o\oplus16\times2e,
 \qquad
 D_h=64+32(3)+16(5)=240.
 \label{eq:app-center-width}
\end{equation}
The flattened widths 60 and 240 count magnetic components of irreducible
tensors; they are not collections of unrelated invariant scalars.

\subsection{Queries, keys, and equivariant values}

At block \(t\), the even scalar channels of the center state provide \(H\)
queries,
\begin{equation}
 \vect q_{ip}^{(t)}
 =W_{p}^{Q,t}\,
 \mathrm{LN}_{h}\!\left(\vect h_{i,\ell=0}^{(t)}\right)
 \in\mathbb R^{d_k},
 \qquad p=1,\ldots,H .
 \label{eq:app-query}
\end{equation}
Only the invariant scalar part of each fixed edge token provides its keys,
\begin{equation}
 \vect k_{ep}^{(t)}
 =W_{p}^{K,t}\,
 \mathrm{LN}_{a}\!\left(\vect a_{e,\ell=0}\right)
 \in\mathbb R^{d_k}.
 \label{eq:app-key}
\end{equation}
The complete edge tensor, including nonscalar irreps, provides one equivariant
value per head,
\begin{equation}
 \vect v_{ep}^{(t)}
 =W_{p}^{V,t}\vect a_e
 \in\mathcal V_h .
 \label{eq:app-value}
\end{equation}
Here \(W_{p}^{V,t}:\mathcal V_a\rightarrow\mathcal V_h\) is an equivariant
linear map. It mixes multiplicity channels only between matching angular
degree and parity and therefore preserves the \(O(3)\) transformation law.
\begin{draftthirtytwoblock}
The operations denoted by \(\mathrm{LN}_{h}\) and \(\mathrm{LN}_{a}\) are
PyTorch affine layer normalizations with \(\epsilon=10^{-5}\), applied over
the multiplicity channels of the even-scalar center and edge sectors,
respectively; they do not mix magnetic components of nonscalar irreps. The
query and key projections are bias-free.
\end{draftthirtytwoblock}
Every block has its own learned \(W^{K,t}\) and \(W^{V,t}\). Thus the keys and
values need not be numerically identical between blocks; ``fixed'' refers to
their dependence on the unchanged token \(\vect a_e\), rather than on a
propagated sender state. More explicitly,
\begin{equation}
 \frac{\partial\vect k_{ep}^{(t)}}{\partial\vect h_{s(e)}^{(t)}}=\vect 0,
 \qquad
 \frac{\partial\vect v_{ep}^{(t)}}{\partial\vect h_{s(e)}^{(t)}}=\vect 0.
 \label{eq:app-fixed-dependency}
\end{equation}
\draftthirtytwo{These are direct computational partial derivatives of one
attention block, with the fixed token and network parameters held constant;
they do not assert that \(\vect h\) and \(\vect a\) have independent
dependence on the underlying coordinates or shared embedding parameters.}
The fixed token does contain the learned species embedding
\(\vect e_{Z_{s(e)}}\); the restriction concerns the absence of a
layer-updated sender state. All reported models use one attention block, while
the implementation permits several blocks with independent projections.

In array notation, these objects have dimensions
\begin{equation}
 \begin{split}
 Q&:\ [N,H,d_k],\qquad
 K:\ [N_{\mathrm e},H,d_k],\\
 \alpha&:\ [N_{\mathrm e},H],\qquad
 V_p:\ [N_{\mathrm e},D_h].
 \end{split}
 \label{eq:app-shapes}
\end{equation}
For example, \(N=40\), \(N_{\mathrm e}=622\), \(H=2\), and \(d_k=32\)
give query and key arrays of shape \([40,2,32]\) and \([622,2,32]\),
respectively, and 622 scores per head.

\subsection{Why the score array is \texorpdfstring{\([N_{\mathrm e},H]\)}
{[Ne,H]}, not \texorpdfstring{\([N_{\mathrm e},N_{\mathrm e},H]\)}
{[Ne,Ne,H]}}

Scaled dot-product attention was introduced in the Transformer as
\(\vect Q\vect K^{\mathsf T}/\sqrt{d_k}\)~\cite{vaswani2017}. This product is
square only for self-attention when the query and key sequences have the same
length. For a particular TRACE center \(i\) and one attention head, collect
its \(n_i\) incoming edge keys and values into
\begin{equation}
 \vect K_i\in\mathbb R^{n_i\times d_k},
 \qquad
 \vect V_i\in\mathbb R^{n_i\times D_h},
 \qquad
 \vect Q_i\in\mathbb R^{1\times d_k}.
 \label{eq:app-local-qkv}
\end{equation}
The ordinary cross-attention score row for that center has shape
\begin{equation}
 \frac{\vect Q_i\vect K_i^{\mathsf T}}{\sqrt{d_k}}
 \in\mathbb R^{1\times n_i}.
 \label{eq:app-local-score}
\end{equation}
TRACE evaluates one such row for every center. Because the \(n_i\) are
different, these ragged rows are stored consecutively in the edge array, with
\begin{equation}
 N_{\mathrm e}=\sum_{i=1}^{N}n_i .
 \label{eq:app-edge-count}
\end{equation}

Equivalently, define a conceptual center--edge mask
\begin{equation}
 M_{ie}=
 \begin{cases}
 1,&r(e)=i,\\
 0,&r(e)\ne i.
 \end{cases}
 \label{eq:app-incidence-mask}
\end{equation}
A dense center--edge tensor would contain \(NN_{\mathrm e}H\) entries, but
only the \(N_{\mathrm e}H\) entries satisfying \(M_{ie}=1\) are valid. The
implementation stores exactly those entries,
\begin{equation}
 \eta_{ep}^{(t)}
 =
 \frac{
 (\vect q_{r(e)p}^{(t)})^{\mathsf T}\vect k_{ep}^{(t)}
 }{\sqrt{d_k}} .
 \label{eq:app-sparse-dot}
\end{equation}
In code this contraction is
\begin{verbatim}
logits = (
    queries[receiver] * keys
).sum(dim=-1) / math.sqrt(key_dim)
\end{verbatim}
The indexing operation gathers the query of the correct receiver for every
edge, changing the query shape from \([N,H,d_k]\) to
\([N_{\mathrm e},H,d_k]\). Elementwise multiplication aligns each valid
query--key pair, and \(\texttt{sum(dim=-1)}\) evaluates
\begin{equation}
 \sum_{\mu=1}^{d_k}q_{r(e),p\mu}k_{e,p\mu}
 =\vect q_{r(e)p}^{\mathsf T}\vect k_{ep}.
 \label{eq:app-code-dot}
\end{equation}
This is the same vector dot product as one entry of
\(\vect Q\vect K^{\mathsf T}\); an explicit transpose is unnecessary because
the valid vectors have already been paired by \(\texttt{receiver}\).

An \([N_{\mathrm e},N_{\mathrm e}]\) score matrix would define edge-to-edge
self-attention, in which every directed edge queries every other edge. Even a
cutoff-local version would require a separate \(n_i\times n_i\) matrix at
each center and \(\sum_i n_i^2\) scores. Neither operation is used in TRACE.
Instead, one environment-conditioned center query selects among the fixed
tokens in its own neighborhood. This relation to attention pooling over an
unordered set is closest at the general level to Set Transformer
\cite{lee2019set}, although TRACE uses a geometry-dependent ACE center query
rather than a learned seed and requires equivariant tensor values. Moreover,
\(\vect q_i^{(t)}\) is formed from the ACE-correlated center state and
therefore depends on all edges in \(\mathcal N(i)\). Each score is consequently
conditioned on the complete local environment even though only one aligned
center--edge dot product is evaluated for each token.

\subsection{Radial logits and cutoff-preserving segment softmax}

The dot product is augmented by invariant radial terms,
\begin{equation}
 s_{ep}^{(t)}
 =
 \eta_{ep}^{(t)}
 +b_p^{(t)}(\vect B(d_e))
 -\mathrm{softplus}(\lambda_p^{(t)})d_e ,
 \label{eq:app-full-logit}
\end{equation}
where the final term is present when the distance penalty is enabled, as it is
for the reported models.
\draftthirtytwo{The implemented radial-bias network is
\(\mathrm{Linear}(N_r,\max(16,4H))\)--SiLU--\(\mathrm{Linear}(\max(16,4H),H)\);
unlike the descriptor radial network, these two linear maps include biases.
The score is dimensionless, so
\(\mathrm{softplus}(\lambda_p^{(t)})\) has reciprocal-distance units in the
chosen coordinate convention.}
The score passed to normalization is
\begin{equation}
 \widetilde s_{ep}^{(t)}
 =\frac{s_{ep}^{(t)}}{\max(T_{\mathrm{att}},10^{-4})}.
 \label{eq:app-temperature}
\end{equation}
The scheduled \(T_{\mathrm{att}}\) is used during training and equals one for
the reported inference calculations. All terms are invariant under \(O(3)\):
the queries and keys are \(0e\) channels, and the remaining functions depend
only on \(d_e\). The score therefore cannot select a preferred spatial
direction.

The softmax is evaluated independently for the incoming edges of every
receiver and every head,
\begin{equation}
 \alpha_{ep}^{(t)}
 =
 \frac{
 f_{\mathrm c}(d_e)\exp(\widetilde s_{ep}^{(t)})
 }{
 1+\displaystyle\sum_{e':r(e')=r(e)}
 f_{\mathrm c}(d_{e'})\exp(\widetilde s_{e'p}^{(t)})
 }.
 \label{eq:app-segment-softmax}
\end{equation}
The unit in the denominator is a null channel with logit zero. Its normalized
weight is
\begin{equation}
 \alpha^{(0,t)}_{ip}
 =
 \frac{1}{
 1+\displaystyle\sum_{e:r(e)=i}
 f_{\mathrm c}(d_e)\exp(\widetilde s^{(t)}_{ep})},
 \qquad
 \alpha^{(0,t)}_{ip}+\sum_{e:r(e)=i}\alpha^{(t)}_{ep}=1 .
 \label{eq:app-null-weight}
\end{equation}
The null channel has no value vector and therefore acts only as a gate on the
magnitude of the neighbor update. In particular, the physical-edge weights
sum to \(1-\alpha^{(0,t)}_{ip}<1\), rather than to one. Without this channel,
if all retained edges shared a small cutoff factor, ordinary softmax
normalization could divide that common factor away. Equation
\eqref{eq:app-segment-softmax} instead gives
\(\alpha_{ep}\rightarrow0\) as \(d_e\rightarrow r_{\mathrm c}\).
For \(f_{\mathrm c}(d_e)>0\), the same normalization can be viewed as an
ordinary softmax over a null logit zero and physical-edge logits
\(\widetilde s_{ep}+\log f_{\mathrm c}(d_e)\); the explicit form avoids taking
\(\log 0\) at the cutoff.

For numerical stability, the implementation uses
\begin{equation}
 m_{ip}=\max\!\left(0,\max_{e:r(e)=i}\widetilde s_{ep}\right),
 \label{eq:app-segment-max}
\end{equation}
followed by
\begin{align}
 \nu_{ep}&=f_{\mathrm c}(d_e)
 \exp(\widetilde s_{ep}-m_{r(e)p}),\\
 \nu_{ip}^{(0)}&=\exp(-m_{ip}),\\
 D_{ip}&=\nu_{ip}^{(0)}+\sum_{e:r(e)=i}\nu_{ep},\qquad
 \alpha_{ep}=\nu_{ep}/D_{r(e)p}.
 \label{eq:app-stable-softmax}
\end{align}
The initial zero in the segment maximum is precisely the null-channel logit.
The maximum and sum are implemented as receiver-indexed scatter reductions.
The exponentials and reductions are explicitly evaluated in float32, the
denominator is bounded below by \(10^{-12}\), and the weights are cast back to
the input type. A center with no incoming edge receives no attention update;
if the complete edge list is empty, the implementation skips attention and
applies only the scalar feed-forward sublayer.

\subsection{Equivariant aggregation, residual update, and scalar feed-forward
map}

For each head, the weighted values are accumulated into their receivers,
\begin{equation}
 \vect z_{ip}^{(t)}
 =\sum_{e:r(e)=i}\alpha_{ep}^{(t)}\vect v_{ep}^{(t)}.
 \label{eq:app-head-sum}
\end{equation}
During training, dropout is applied to \(\alpha_{ep}\) after normalization and
before Eq.~\eqref{eq:app-head-sum}; validation and inference use the
undropped weights. The reported configurations use dropout probability
0.03. If \(m_{ep}\) is an independent Bernoulli mask with retention
probability \(1-p_{\mathrm d}\), the coefficient used during training is
\begin{equation}
 \widehat\alpha_{ep}
 =\frac{m_{ep}}{1-p_{\mathrm d}}\alpha_{ep}.
 \label{eq:app-attention-dropout}
\end{equation}
This is the standard inverted-dropout convention, so
\(\mathbb E[\widehat\alpha_{ep}]=\alpha_{ep}\). Dropout is not followed by a
second normalization.

The active scatter operation is schematically
\begin{verbatim}
update.index_add_(
    0, receiver, alpha[:, p:p+1] * values
)
\end{verbatim}
This receiver-only accumulation is required by the directed-edge convention.
Token \(j\rightarrow i\) describes sender \(j\) inside the environment whose
state is stored at receiver \(i\), so it contributes to \(\vect h_i\). In a
full directed list, the reverse token \(i\rightarrow j\) separately
contributes to \(\vect h_j\). The sender index has already entered the species
embedding and geometry in Eq.~\eqref{eq:app-edge-token}; it is intentionally
not used to fetch a current sender hidden state.

The head sum is averaged, projected equivariantly, and added through an
irrep-wise residual scale,
\begin{equation}
 \widetilde{\vect h}_i^{(t)}
 =
 \vect h_i^{(t)}
 +\vect\gamma_{\mathrm{attn}}^{(t)}\odot
 W^{O,t}\!\left(\frac{1}{H}\sum_{p=1}^{H}\vect z_{ip}^{(t)}\right).
 \label{eq:app-attention-residual}
\end{equation}
Thus TRACE averages head updates in their common equivariant output space; it
does not concatenate the heads as in the original sequence Transformer.
\begin{draftthirtytwoblock}
Equation~\eqref{eq:app-cutoff-order} and the linear value map give
\(\vect v_{ep}=\mathcal O(f_{\mathrm c})\), while \(\alpha_{ep}\) contains an
explicit cutoff factor. Each edge contribution to the attention update is
therefore \(\mathcal O(f_{\mathrm c}^{\,2})\) as
\(d_e\rightarrow r_{\mathrm c}^{-}\). With the quintic envelope this is
\(\mathcal O[(r_{\mathrm c}-d_e)^6]\), so the contribution and its first two
radial derivatives vanish at the cutoff.

For checkpoint compatibility, the implementation stores component parameters
\(\theta_{c m}^{(\ell,t)}\), but uses only their mean within each irrep copy,
\begin{equation}
 \gamma_{c}^{(\ell,t)}
 =\frac{1}{2\ell+1}\sum_{m=-\ell}^{\ell}
 \theta_{c m}^{(\ell,t)} ,
 \qquad
 [\vect\gamma_{\mathrm{attn}}^{(t)}\odot
 \vect x]_{c m}^{(\ell)}
 =\gamma_{c}^{(\ell,t)}x_{c m}^{(\ell)} .
 \label{eq:app-irrep-scale}
\end{equation}
Thus the effective scale is one scalar per irrep copy, broadcast over all
magnetic components. The reported models initialize every stored component
to \(10^{-2}\). Because \(\alpha_{ep}\) and \(\gamma_c^{(\ell,t)}\) are
invariant and \(\vect v_{ep}\) is equivariant, each \(\vect z_{ip}\) and the
residual update transform equivariantly.
\end{draftthirtytwoblock}

The feed-forward sublayer does not apply an unconstrained componentwise MLP to
nonscalar tensors. It concatenates the even scalar channels with invariant
squared norms
\begin{equation}
 n_{ic\ell}^{(t)}
 =\left\lVert\widetilde{\vect h}_{ic}^{(t,\ell)}\right\rVert^2
 =\sum_{m=-\ell}^{\ell}
 \left|\widetilde h_{icm}^{(t,\ell)}\right|^2,
 \qquad \ell>0,
 \label{eq:app-tensor-norm}
\end{equation}
\begin{draftthirtytwoblock}
The exact scalar path is
\begin{align}
 \vect\xi_i^{(t)}
 &=\mathrm{LN}_{s}\!\left[
 \widetilde{\vect h}_{i,\ell=0}^{(t)},
 \{n_{ic\ell}^{(t)}\}_{\ell>0}\right],\\
 \vect g_i^{(t)}
 &=\mathrm{Drop}_{p_{\mathrm d}}\!\left[
 \mathrm{SiLU}\!\left(W_1^{(t)}\vect\xi_i^{(t)}+\vect b_1^{(t)}\right)
 \right],\\
 \Delta\vect s_i^{(t)}
 &=\mathrm{Drop}_{p_{\mathrm d}}\!\left[
 W_2^{(t)}\vect g_i^{(t)}+\vect b_2^{(t)}
 \right],\\
 \vect h_{i,\ell=0}^{(t+1)}
 &=\widetilde{\vect h}_{i,\ell=0}^{(t)}
 +\vect\gamma_{\mathrm{FFN}}^{(t)}
 \odot\Delta\vect s_i^{(t)},\\
 \vect h_{i,\ell>0}^{(t+1)}
 &=\widetilde{\vect h}_{i,\ell>0}^{(t)} .
 \label{eq:app-scalar-ffn}
\end{align}
Here \(\mathrm{Drop}_{p_{\mathrm d}}\) is inverted dropout during training and
the identity during validation and inference. The reported models use
\(p_{\mathrm d}=0.03\) and initialize
\(\vect\gamma_{\mathrm{FFN}}^{(t)}\) to \(10^{-2}\). For the illustrative
representation in Eqs.~\eqref{eq:app-edge-width} and
\eqref{eq:app-center-width}, \(\vect\xi_i\) has
\(64+32+16=112\) components and the default hidden width is \(2(64)=128\).
\(\mathrm{LN}_{s}\) is an affine layer normalization with
\(\epsilon=10^{-5}\).
\end{draftthirtytwoblock}
The nonscalar tensors remain unchanged in this sublayer. An invariant atomic
readout then gives \(E_i\). The coordinate- and cell-independent reference
\(E_{\mathrm{ref}}\) is added by the training and calculator wrappers, so
\(E=\sum_iE_i+E_{\mathrm{ref}}\) yields the same conservative force and stress
derivatives as the learned residual energy.

\subsection{Algorithmic sequence}

For one block, the implemented calculation can be summarized as follows:
\begin{enumerate}
 \item Normalize the current center representation and project its even scalar
 channels to \(Q[N,H,d_k]\).
 \item Normalize the scalar part of the fixed edge tensor and project it to
 \(K[N_{\mathrm e},H,d_k]\).
 \item Gather \(\texttt{Q[receiver]}\), contract the aligned query--key pairs over
 \(d_k\), and add the radial bias and nonnegative distance penalty.
 \item Apply the lower-bounded attention-temperature factor and the
 cutoff-preserving segment softmax of Eq.~\eqref{eq:app-stable-softmax};
 during training, apply post-softmax dropout without renormalization.
 \item For each head, map the complete fixed edge tensor equivariantly to its
 value, multiply by the scalar attention weight, and scatter-add the result to
 the receiver.
 \item Average the heads, apply the equivariant output projection and tied
 irrep-wise layer scale, and add the attention residual.
 \item \draftthirtytwo{Form the invariant tensor norms, normalize the combined
 scalar-and-norm vector, apply
 linear--SiLU--dropout--linear--dropout, multiply the scalar residual by its
 layer-scale vector, and add it only to the scalar channels.}
\end{enumerate}
At no step is \(\vect h_{s(e)}^{(t)}\) read to construct the key or value.
Thus increasing the number of blocks changes the nonlinear interrogation of
one fixed environment but does not propagate an updated state through a chain
of atoms.

\subsection{Permutation symmetry, locality, and linear scaling}

Permuting edge storage leaves receiver-indexed reductions unchanged. Relabeling
equivalent atoms correspondingly relabels the center outputs, while their
energy sum restores permutation invariance. Because the attention weights are
\(O(3)\)-invariant scalars, the values are equivariant, parities are explicit,
and the readout is restricted to \(0e\), the energy is invariant under
\(O(3)\).

The sparsity comes from two separate restrictions. First, the physical
neighbor list retains only periodic directed edges inside \(r_{\mathrm c}\).
Second, center \(i\) attends only to tokens whose receiver is \(i\). For mean
neighbor count \(\overline n\),
\begin{equation}
 N_{\mathrm e}\simeq N\overline n .
 \label{eq:app-linear-edges}
\end{equation}
Every retained geometric edge is used; there is no learned top-\(k\)
selection, thresholding, or stochastic pruning. Query formation costs
\(\mathcal O(NHd_k)\); the aligned query--key contraction costs
\(\mathcal O(N_{\mathrm e}Hd_k)\) and stores
\(\mathcal O(N_{\mathrm e}H)\) logits. At fixed widths, edge maps and
reductions cost \(\mathcal O(N_{\mathrm e})\), center correlations and
readout cost \(\mathcal O(N)\), and activation memory is
\(\mathcal O(N+N_{\mathrm e})\). The model is therefore linear in system size
for bounded \(\overline n\). Global atom and edge self-attention instead store
\(\mathcal O(N^2H)\) and \(\mathcal O(N_{\mathrm e}^2H)\) scores,
respectively; local edge-to-edge self-attention stores
\(\mathcal O(H\sum_i n_i^2)\).

Linear asymptotic cost is not a hardware benchmark: throughput also depends
on device-resident neighbor lists, fused kernels, compiled derivatives,
memory traffic, force and virial accumulation, and inter-rank communication.
Locality fixes the \(r_{\mathrm c}\) ghost halo and removes hidden-state
exchange, but not these costs.

\end{draftthirtyoneblock}

\end{document}